\documentclass[a4paper,fleqn]{cas-sc}

\newcounter{experiment} 

\usepackage{amssymb}
\usepackage{verbatim}
\usepackage{multirow}
\usepackage{array}
\usepackage{longtable}
\usepackage{booktabs}
\usepackage{tabularray}
\usepackage{hyperref}
\usepackage{caption}
\usepackage{tabularx}
\usepackage{graphicx}
\usepackage{multicol}
\usepackage{geometry}
\usepackage{url}
\usepackage{lscape}
\usepackage{pdflscape}
\usepackage{xcolor}
\usepackage{float}
\usepackage{comment}
\usepackage{tcolorbox}
\usepackage{amsmath}

\usepackage{soul}
\usepackage{hyperref}
\usepackage{multirow}
\usepackage{listings}
\usepackage{fancybox}
\usepackage{enumitem}
\usepackage{graphicx}
\usepackage{color}
\usepackage{array}
\usepackage{amsmath}
\usepackage{xspace}
\usepackage{url}
\usepackage{tikz}
\usepackage{caption}
\usepackage{pgfplots}
\usepackage{balance}
\usepackage{subcaption}
\pgfplotsset{compat=1.16}
\usepackage{pgf-pie}
\usetikzlibrary{pgfplots.statistics,calc}
\usepackage{algorithm}
\usepackage{algorithmic}
\usepackage{adjustbox}
\usepackage{makecell}

\usepackage{tcolorbox}
\makeatletter
\newcommand{\mybox}[1]{%
	\setbox0=\hbox{#1}%
	\setlength{\@tempdima}{\dimexpr\wd0+13pt}%
	\begin{tcolorbox}[boxrule=0.5pt, colback=white, arc=4pt,
		left=6pt,right=6pt,top=6pt,bottom=6pt,boxsep=0pt]
		#1
	\end{tcolorbox}
}

\definecolor{songcolor}{RGB}{191,191,191}
\newcommand{\todoc}[2]{{\textcolor{#1}{\textbf{#2}}}}
\newcommand{\todored}[1]{\todoc{red}{\textbf{[[#1]]}}}

\newcommand{\alvine}[1]{\todored{Alvine: #1}}

\usepackage[authoryear]{natbib}

\def\tsc#1{\csdef{#1}{\textsc{\lowercase{#1}}\xspace}}
\tsc{WGM}
\tsc{QE}

\begin{document}
\let\WriteBookmarks\relax
\def\floatpagepagefraction{1}
\def\textpagefraction{.001}

\shortauthors{Odu et al.}

\title [mode = title]{Automatic Instantiation of Assurance Cases from Patterns Using Large Language Models}  

\shorttitle{Automatic Instantiation of Assurance Cases from Patterns Using Large Language Models}

\fntext[1]{}


\author[inst1]{Oluwafemi Odu}

\affiliation[inst1]{organization={Lassonde School Of Engineering},
            addressline={York University}, 
            city={Toronto},
            country={Canada}}

\author[inst1]{Alvine B. Belle}
\author[inst1]{Song Wang}

\author[inst2]{Segla Kpodjedo}

\author[inst3]{Timothy C. Lethbridge}

\author[inst1]{Hadi Hemmati}

\affiliation[inst2]{organization={Department of Software Engineering and Information Technology},
            addressline={École de technologie supérieure}, 
            city={Montreal},
            country={Canada}}

            \affiliation[inst3]{organization={School of Electrical Engineering and Computer Science},
            addressline={University of Ottawa}, 
            city={Ottawa},
            country={Canada}}

\newcommand{\segla}[1]{\textcolor{red}{[Segla: #1]}}
\newcommand{\timlethbridge}[1]{\textcolor{orange}{[TimLethbridge: #1]}}
\newcommand{\Femi}[1]{\textcolor{Blue}{[Femi: #1]}}

\begin{abstract}
An assurance case is a structured set of arguments supported by evidence, demonstrating that a system’s non-functional requirements (e.g., safety, security, reliability) have been correctly implemented.
Assurance case patterns serve as templates derived from previous successful assurance cases, aimed at facilitating the creation of new assurance cases. Despite the use of these patterns to generate assurance cases, their instantiation remains a largely manual and error-prone process that heavily relies on domain expertise. Thus, exploring techniques to support their automatic instantiation becomes crucial. This study aims to investigate the potential of Large Language Models (LLMs) in automating the generation of assurance cases that comply with specific patterns. Specifically, we formalize assurance case patterns using predicate-based rules and then utilize LLMs, i.e., GPT-4o and GPT-4 Turbo, to automatically instantiate assurance cases from these formalized patterns. Our findings suggest that LLMs can generate assurance cases that comply with the given patterns. However, this study also highlights that LLMs may struggle with understanding some nuances related to pattern-specific relationships. While LLMs exhibit potential in the automatic generation of assurance cases, their capabilities still fall short compared to human experts. Therefore, a semi-automatic approach to instantiating assurance cases may be more practical at this time. 

\end{abstract}


\maketitle

\section{Introduction}
Complex critical systems such as cyber-physical systems, are increasingly designed to be interoperable and interconnected. The growing complexity of their configurations and operations in dynamic environments underscores the importance of system assurance. Ensuring the correct implementation of non-functional requirements, such as safety and security, is vital to prevent these systems' failure that could result in severe consequences, including fatalities and financial losses \citep{b83,b86}. 

Assurance cases (ACs) are structured arguments with a supporting body of evidence that allow demonstrating that the non-functional requirements of a system have been correctly implemented \citep{b7}. Assurance cases are utilized across various domains
(e.g., medicine \citep{b8,b16,b17}, automotive \citep{b18,b19,b20} to support the certification and compliance of critical systems with industry standards (e.g., ISO 26262 \citep{b84}, DO-178C \citep{b85}). 
Manually creating assurance cases can be time-consuming, especially for large, complex, and interconnected systems \citep{b94, b107}. For instance, \citet{b94} noted that an assurance case for an air traffic control system may consist of over 500 pages and include references to 400 documents. This suggests that the manual creation and subsequent modifications of an initial assurance case draft can take several months \citep{b139}.  To facilitate this creation process, practitioners use assurance case patterns (ACPs). These patterns are templates composed of evidence-based arguments derived from previous successful assurance cases. Practitioners instantiate assurance case patterns with system-specific information to create new assurance cases more efficiently. Several notations allow representing ACs and ACPs. These include the Goal Structuring Notation (GSN) \citep{b22} and the Claims-Arguments-Evidence (CAE) \citep{b7} notation.

Despite the use of assurance case patterns to create assurance cases, the instantiation process remains tedious, error-prone, and time-consuming. This is primarily due to the heterogeneous nature of system artifacts and the complexity of mission-critical systems \citep{b88}. Instantiating these patterns with system-specific information still requires domain expertise to efficiently extract the necessary system artifacts. Experts must then manually replace the abstract elements within these patterns with concrete values from the extracted artifacts to create an assurance case that complies with the given pattern(s).

There is a wealth of literature on the automatic instantiation of assurance case patterns. However, it is worth noting that in this literature, the expressions \textit{'instantiation of assurance case patterns to generate assurance cases'} and \textit{'instantiation of assurance cases from patterns'} are frequently used interchangeably. Most of the existing approaches for instantiating assurance cases from patterns strongly depend on the model based engineering approach that support extraction of information from system models (e.g., model-based design \citep{b88, b100, b101, b102, b103}. 
However, a strong dependence on model based engineering approach can limit the application of assurance case patterns in creating assurance cases for systems that do not conform to this design approach. 
 This highlights the need for new techniques to automatically instantiate assurance cases from patterns for any given system, irrespective of its design methodology.

The rapid adoption of generative AI technologies like OpenAI's GPT series has fostered the automatic  generation of content and spurred their increasing use in the automation of several software engineering tasks \citep{b104}. To capitalize on this momentum, we propose a novel approach that utilizes Large Language Models (LLMs) to automatically instantiate assurance cases complying with a specified assurance case pattern(s). 
Our experiment results reveal LLMs can effectively generate assurance cases. Thus, our contributions are fourfold:
\begin{itemize}
    \item We propose a novel method for formalizing ACPs  into predicate-based rules complying with  GSN. This allows for capturing the internal structure of ACPs more generically and uniformly.
    \item We explore the use of LLMs to automatically generate assurance cases complying with formalized ACPs. 
    \item We experiment with two popular and very recent LLMs (i.e. GPT-4 Turbo and GPT-4o) to explore their ability to automatically generate ACs from patterns. 
    \item We release the dataset and source code of our experiments to help other researchers replicate and extend our study\footnote{\url{https://doi.org/10.6084/m9.figshare.27103225.v2}}.
\end{itemize}

The remainder of this paper is organized as follows: Section~\ref{sec:2} presents some background concepts. Section \ref{sec:RelatedWork} discusses related work. Section \ref{sec:approach} presents our methodology. Section \ref{sec:Setup} describes the experimental setup. Section \ref{sec:Result} reports the results of our study. In Section \ref{sec:Discussion}, we discuss our results. Section \ref{sec:Validity_Threats} identifies the threats to validity associated with our study. We conclude and outline future work in Section \ref{sec:Conclusion}.


\section{Background}
\label{sec:2}
\subsection{Assurance Case}

An assurance case is a well-established, structured, reasoned, and auditable set of arguments designed to support a specific goal \citep{b6}. These arguments are often supported by evidence demonstrating that a system meets desirable non-functional requirements (e.g., safety, security). There are several types of assurance cases, each focusing on a specific non-functional requirement: safety cases \citep{b8, b9}, security cases \citep{b10, b11}, 
and reliability cases \citep{b13}.

Assurance cases are utilized to prevent system failure in various domains, including medicine \citep{b8,b16,b17} and automotive \citep{b18,b19,b20}. They are also used to ensure the reliability of mission-critical systems and facilitate certification in line with industry standards (e.g., ISO 26262, DO-178C). Regulatory bodies like the Food and Drug Administration (FDA) advocate for the use of assurance cases to bolster the safety confidence of medical devices during their approval process \citep{b10}.

An assurance case comprises three primary components \citep{b23, b136}: (1) a top claim (root claim) which is usually subdivided into sub-claims. This top claim serves as the fundamental statement indicating that the system fulfills a specific requirement. (2) body of evidence supporting both the sub-claims and the root claim. (3) a collection of structured arguments that establish connections between the evidence and the sub-claims, linking all sub-claims to the top claim of the assurance case \citep{b21}.


\subsection{Assurance Case Pattern}

Similar to design patterns used in software engineering (SE), assurance case patterns are templates formed from common repeated structures and previous successful assurance cases \citep{b1}. These assurance case patterns contain placeholders filled with generic information that can be replaced with system-specific information during their instantiation \citep{b88, b1}. The use of assurance case patterns fosters the reuse and eases the creation of assurance cases. There are various types of assurance case patterns based on the non-functional requirements they target. Assurance case patterns are also used to mitigate assurance deficits \citep{b27, b29, b133}. Assurance deficits refer to \textit{"any knowledge gap that prohibits perfect confidence"} in an assurance case \citep{b25}.



\subsection{Representation of Assurance Cases and Assurance Case Patterns}

\subsubsection{Assurance Case Representation}
\label{sec:AC_Desc}
\begin{figure} [!h]
    \centering
    \includegraphics[width=1\linewidth]{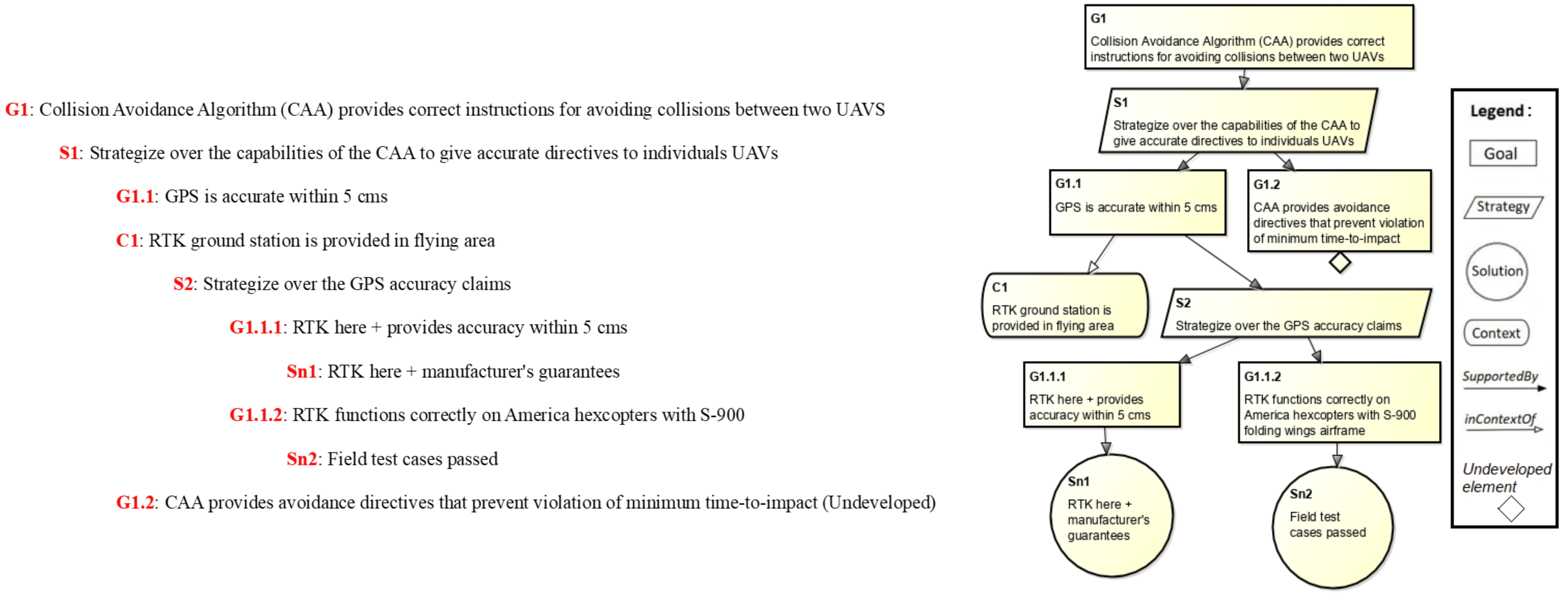}
    \caption{On the right, an example of a partial safety case (GSN diagram) adapted from  \citet{b106}; on the left, the equivalent of the safety case in  structured prose}
    \label{fig:AC-example}
\end{figure}

Several notations allow for representing assurance cases and can be broadly categorized into textual and graphical notations. 
\citet{b90} described five text-based notations for representing assurance cases, including normal (i.e. unstructured) prose representation and structured prose representation. The latter is a notation that introduces a structured format to address the typical verbosity, lack of structure, and ambiguity characterizing the normal prose representations of assurance cases \citep{b89, b91}. 

Graphical notations also address the limitations of unstructured representations by improving the clarity and structure of assurance cases.
Graphical notations include GSN (Goal structuring Notation) \citep{b22}, CAE (Claim-Argument-Evidence) \citep{b7}, and Eliminative Argumentation (EA) \citep{b24}. The Object Management Group (OMG) recently introduced SACM (Structured Assurance Case Metamodel) \citep{b23} to promote interoperability and standardization \citep{b3}. SACM aligns with existing assurance case notations (e.g., GSN, CAE). Still, GSN is the most popular notation \citep{b3}. GSN supports the representation of an assurance case as a \textit{goal structure}. The latter is a GSN diagram depicted as a tree-like structure. The GSN  standard \citep{b22} proposes the following six main GSN elements to represent assurance cases: 

\begin{itemize}
  \item \textbf{A Goal} is depicted as a rectangle and represents the main claim or a sub-claim. Examples are G1 through G4 in Figure \ref{fig:ACP-example}.
  \item \textbf{A Strategy} is depicted as a parallelogram and describes the inference between a goal and its sub-goals. See S1 in Figure \ref{fig:ACP-example}.
  \item \textbf{A Solution} is depicted as a circle and represents the evidence supporting an argument or goal. See Sn1 and Sn2 in Figure \ref{fig:AC-example}. 
  \item  \textbf{A Context} is rendered as a rounded rectangle, presenting a contextual artifact. This can be a reference to contextual information, or a statement. See C1 in Figure \ref{fig:ACP-example}.
  \item  \textbf{An Assumption} is rendered as an ellipse with the letter `A' at the top or the bottom right denoting an intentionally unsubstantiated statement.
  \item \textbf{A Justification} is rendered as an ellipse with the letter `J' at the top or the bottom right and presents a statement of rationale for the inclusion or wording of a GSN element.
 \end{itemize}
 
Assurance cases' claims, evidence, and arguments respectively map to GSN goals, solutions, and strategies \citep{b136}. It is possible to decorate GSN elements using the \textit{Undeveloped} decorator. The latter allows indicating a GSN element has not been developed yet \citep{b22}. It is depicted as a hollow diamond applied to the bottom center of an element \citep{b22}.  Furthermore, the GSN  standard \citep{b22} defines two main relationships between GSN elements: \textit{SupportedBy} and \textit{InContextOf}. 
 \textit{SupportedBy} is depicted as a line with a solid arrowhead and represents supporting relationships between GSN elements. \textit{InContextOf} is depicted as a line with a hollow arrowhead and represents a contextual relationship between GSN elements.
 
 The GSN standard \citep{b22} proposes guidelines to convert an assurance case represented in the textual format (eg., structured prose) into a GSN diagram. Figure  \ref{fig:AC-example} shows an excerpt of an assurance case adapted from \cite{b106}. This excerpt is represented in the GSN and in the structured prose.

\subsubsection{Assurance Case Pattern Representation}

GSN and some reference literature on GSN patterns (e.g., \cite{b31, b87, b128}) also propose the following additional decorators to help  represent assurance case patterns:

 \begin{figure} [!h]
    \centering
    \includegraphics[width=0.85\linewidth]{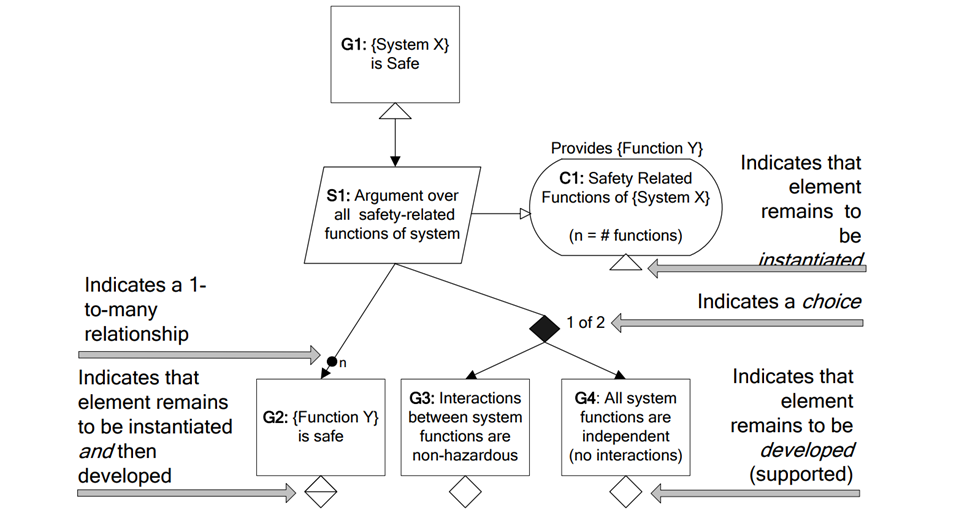}
    \caption{A sample assurance case pattern adapted from \citet{b32}}
    \label{fig:ACP-example}
\end{figure}

\begin{itemize}
    
\item  \textbf{Uninstantiated} - This decorator is depicted with a small triangle applied to the bottom center of an element. It allows indicating that a GSN element is yet to be instantiated, i.e., an abstract element in a placeholder needs to be replaced by a concrete instance~\citep{b22}.
\item \textbf{Undeveloped and Uninstantiated} - Both the \textit{undeveloped} and \textit{uninstantiated} decorators are overlaid to form this decorator. It denotes that a GSN element requires both further development and instantiation.
\item \textbf{Parameterized expressions within Placeholders} - Parameterized expressions are abstract expressions inside placeholders that need to be replaced with concrete information \citep{b31, b87, b128}.

\item \textbf{Multiplicity} - Multiplicity symbols allow describing how many instances of one element type relate to another element. These symbols are generalized n-ary relationships between GSN element  \citep{b22}.
\item \textbf{Optionality} - It represents optional and alternative relationships between GSN elements which generalizes n-of-m choices between GSN elements  \citep{b22}.
\item \textbf{Choice}. This decorator is depicted as a solid diamond. The choice decorator denotes possible alternatives in satisfying a relationship  \citep{b22}. 
\end{itemize}

Figure \ref{fig:ACP-example} shows a sample safety case pattern adapted from \cite{b32} and represented using GSN.

\subsection{Large Language Models (LLMs)}

LLMs are advanced artificial intelligence systems with computational ability to generate human language, with at least the appearance of understanding it \citep{b92,b60}. These models are complex neural network structures with massive parameter sizes and trained on vast amounts of data from diverse sources \citep{b60}.  LLMs can generate new content, answer questions, and capture inherent rules of a domain \citep{b92, b68, b124}. These capabilities of LLMs have ensured their wide application across various fields including automated software engineering. 
 Some examples of LLMs include the GPT series by OpenAI \citep{b104}, BERT \citep{b62}, T5 \citep{b63}, and ERNIE \citep{b64}.

Prompt engineering refers to the different techniques used to provide instructions and guidelines to an LLM to ensure a desired generated response \citep{b65}. It improves the efficiency and quality of LLMs responses. The most popular prompting techniques include the Chain-of-Thought (CoT) prompting technique \citep{b66}, Zero-shot prompting, \citep{b69}, and Few-shot prompting \citep{b70}.  
CoT utilizes a series of intermediate reasoning steps to significantly improve the ability of LLMs to perform complex reasoning tasks \citep{b66}. 
Zero-shot prompting \citep{b69} — a technique whereby we prompt an LLM without any examples, attempting to take advantage of the reasoning patterns it has extracted. Few-shot prompting \citep{b70} is a technique whereby we prompt an LLM with several concrete examples of task performance so that the model can learn from these examples. 

Rule distillation \citep{b68,b67} is a technique enabling LLMs to learn and acquire knowledge from rules or instructions. 
It involves extracting knowledge from defined textual rules and then explicitly encoding this knowledge into LLM parameters. This  ensures that LLMs can effectively comprehend and apply the distilled rules \citep{b68,b67}.

\section{Related Work}
\label{sec:RelatedWork}
\subsection{Formalization of Assurance Cases and Assurance Case Patterns}

\citet{b39} proposed a formal definition of an assurance case pattern as a tuple characterized by a directed hypergraph with specific labeling functions to enhance pattern utilization. \citet{b31} introduced an assurance case language based on GSN, validated through the D-Case Editor tool, which supports GSN patterns and modules. Expanding on the concept of assurance case languages, \citet{b71} developed \textit{`CyberGSN'}, integrating informal GSN elements with formal Cyberlogic to facilitate safety case creation and maintenance. To address semantic correctness and logical consistency in assurance cases, \citet{b95} developed a framework that converts assurance cases into Prolog predicates and utilizes Constraint Answer Set Programming (CASP) to ensure consistency and completeness of the arguments and evidence in assurance cases.

To assess the benefits associated with the formalization of assurance arguments about a system property, \citet{b40} conducted a survey of twenty studies focusing on proposed formal assurance arguments. Their result revealed that majority of these studies speculate on the advantages of formalism without presenting concrete proof to substantiate these presumed advantages.

Our work is similar to  \citet{b114}, in which the authors extracted predicates from the structural rules embedded in EA. They then used these predicates to create predicate-based rules that they incorporated into GPT-4 Turbo prompts to investigate the effectiveness of that LLM in identifying defeaters within assurance cases represented in EA notation. Defeaters refer to \textit{``arguments that can undermine the effectiveness of assurance cases by compromising the reliability and adequacy of these assurance cases in verifying a system's capabilities such as safety and security''} \citet{b114}. In contrast, our work introduces a novel pattern formalization method that uses formal predicates to capture the internal structure and relationships among elements within an assurance case pattern presented in GSN. By leveraging these formal predicates,  we enable LLMs including GPT-4 Turbo and GPT-4o, to automatically generate assurance cases that adhere to the specified formalized assurance case pattern(s).

Our approach focuses on GSN and utilizes a formalized pattern to systematically generate assurance cases. This key distinction sets our work apart from previous research, which did not use formalized assurance case patterns for the creation of assurance cases and did not assess the performance of various LLMs in generating these assurance cases.

\subsection{Automatic instantiation of assurance case patterns}
\label{sec:pattern_automation}

Several approaches support the automatic instantiation of assurance case patterns (e.g., \citep{b98, b99, b122, b123}). For instance, some approaches (e.g., \citep{b98, b99}) used a weaving method with the model-based engineering approach for instantiating assurance case patterns. This method weaves assurance case patterns with system models, facilitating the extraction of system-specific information or artifacts from system models and mapping these artifacts to placeholders in the ACP to generate an assurance case. Other approaches (e.g., \citep{b122, b123}) utilized reference tables to keep track of system artifacts, requirements, and the mapping of these artifacts to placeholders in the ACP. However, these approaches strongly depend on specific engineering methodologies that focus on extracting information from system models (e.g., model-based design \citep{b88, b100, b101, b102, b103}). 
Additionally, the use and maintenance of reference tables can be complex and challenging, especially for large systems with numerous requirements and artifacts as evidence. This complexity can lead to the omission of important artifacts or evidence if the reference table is not well maintained. Therefore, it is crucial to devise new techniques to automatically instantiate assurance case patterns and create assurance cases for systems regardless of their design methodology. 

\subsection{Rule-based Learning in LLMs}

LLMs sometimes produce inaccurate results or \textit{'hallucinate'} \citep{b72}. To address this issue, recent research \citep{b68, b67} recommended integrating rule-based knowledge into LLMs. This allows LLMs to rely on structured rules when there are insufficient example-based learning resources, thereby enhancing their accuracy and reliability. 
\citet{b68} presented a novel learning paradigm allowing LLMs to assimilate knowledge from rules in a manner akin to human learning processes. Their approach leverages the LLMs' in-context capabilities to first extract knowledge from textual rules and then encode this rule knowledge explicitly by training the model using in-context signals.  
\citet{b73} introduced a novel `grammar prompting' technique to improve the ability of LLMs for generating strings from structured languages using a domain-specific grammar in Backus-Naur Form (BNF). This method augments each example with a specialized grammar sufficient for the output, and the LLM predicts a BNF grammar from the input to generate the output accordingly. Their experiments show that this approach enables LLMs to effectively handle a variety of domain-specific language generation tasks, including semantic parsing and molecule generation.

\subsection{LLMs for Software Modeling}

LLMs are currently being utilized for a variety of downstream SE tasks, including software defect prediction \citep{b33}, static code analysis \citep{b132}, automated program repair \citep{b34}, code generation \citep{b35}, and software modeling \citep{b74,b75}. In the field of software modeling, \citet{b74}  investigated the use of LLMs, specifically GPT-3.5 and GPT-4, to fully automate domain modeling. 
They concluded that including examples in prompts significantly improves the performance of LLMs. Also, \citet{b75} applied GPT-4 in goal-oriented modeling, focusing on its use with the Goal-oriented Requirement Language (GRL). Their results showed that GPT-4 can generate basic goal models, though its outputs often require manual domain-specific adjustments and validation. 
\citet{b96} utilized few-shot prompt learning to ease the completion of domain diagrams (eg., UML class and activity diagrams) without requiring extensive training data. They used semantic mappings to convert modeling formalisms into meaningful patterns of tokens that LLMs can understand 
to improve and complete modeling activities.  
\citet{b107} conducted an evaluation of GPT-4's proficiency in understanding and generating GSN elements and safety cases. They extracted intricate structural and syntactic rules from the GSN standard to formulate 19 evaluative questions divided into rule-based and generation-based questions. They assessed GPT-4's comprehension of these rules and its capability to generate GSN elements. Additionally, using both contextual and domain information, they performed experiments to evaluate GPT-4's ability in producing safety cases that are structurally, semantically, and reasonably accurate.  
\citet{b97} analyzed EA reference documents to extract both the structural and semantic rules that EA embodies. These rules served as the foundation for crafting the EA-based questions they used to evaluate GPT-4 Turbo's proficiency in understanding EA as well as its ability to generate EA elements such as defeaters. Unlike our approach, both \citet{b107} and \citet{b97} did not rely on formalized assurance case patterns to guide the generation of assurance cases and did not compare the performance of various LLMs in generating assurance cases.

\section{Approach}
\label{sec:approach}

\begin{figure} [t!]
    \centering
    \includegraphics[width=0.85\linewidth]{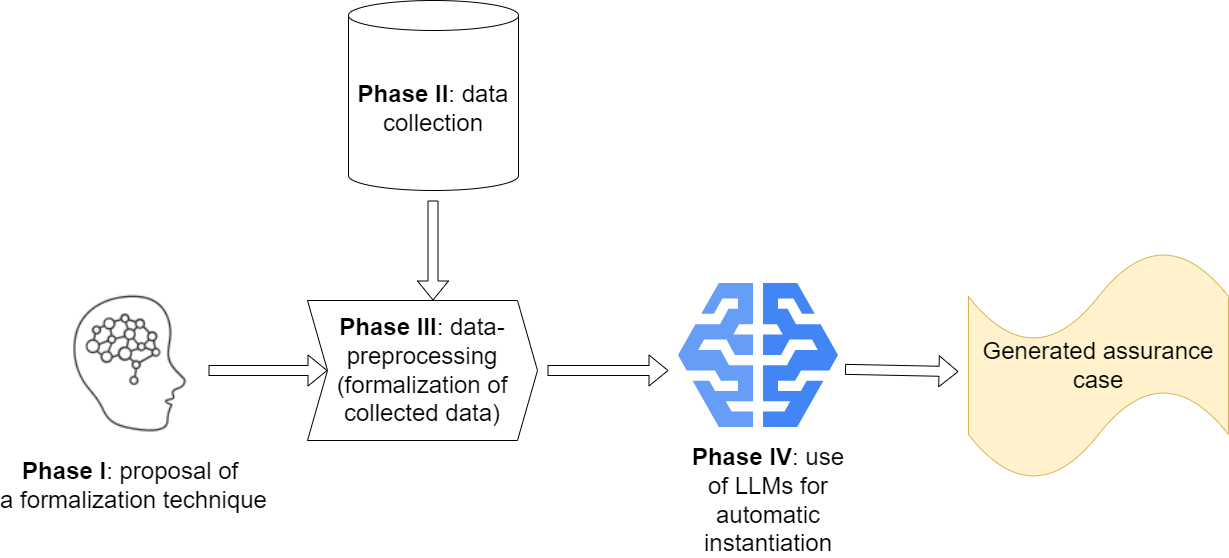}
    \caption{High-level overview of our approach}
    \label{fig:approach-overview}
\end{figure}

The novelty of our approach lies in the use of LLMs to guide the automatic generation of assurance cases from formalized assurance case patterns. This ensures that the LLMs at hand leverage recurring argumentation structures (i.e. patterns) to guide the generation of assurance cases. 
Figure  \ref{fig:approach-overview} shows a high-level overview of our approach. That approach consists of four phases that we describe below.

\subsection{Phase I: Formalization of Assurance Case Patterns into Predicates}
\label{sec:Formalization}

Inspired by the foundational work of \citet{b39} which formalizes ACs and ACPs, we propose a set of predicate-based rules to represent and therefore formalize assurance cases and assurance case patterns. The use of predicates for formalization allows capturing the properties and relationships among the elements of an assurance case and assurance case pattern \citep{b95}. 

To enhance the usability and understanding of LLMs within graphical notations like GSN, we propose integrating rule-based knowledge into LLMs using our predicate-based rules. For this purpose, we first create predicates allowing us to formalize assurance patterns as a set of predicates rendered in a textual format complying with the very popular GSN. 
The predicate-based format is suitable for LLM ingestion and can be considered as an advanced and more formal structured prose. 
We construct our predicates based on the guidelines, elements, relationships, and decorators that the GSN  Standard \citep{b22} outlines. Thus, we formulate a predicate for each GSN core element, relationship, and decorator represented in GSN. This results in three categories of predicates that we further discuss below. 

\subsubsection{Predicate for Formalizing an Assurance Case and Its Decorators}
We propose the following predicates to formalize an assurance case complying with GSN:
    
    \begin{itemize}
            \item \textbf{Goal (G)}: True if G is a goal within the assurance case. This predicate can be represented as Goal (ID, Description) where ID is the unique identifier for the goal, description is the textual description of the goal.
           
             \item \textbf{Strategy (S)}: True if S is a strategy within the assurance case. This predicate can be represented as Strategy (ID, Description) where ID is the unique identifier for the strategy and description is the textual description of the strategy.
             
             \item \textbf{Solution (Sn)}: True if Sn is a piece of evidence within the assurance case. This predicate can be represented as Solution (ID, Description) where ID is the unique identifier for the evidence and description is the textual description of the evidence.
             
             \item \textbf{Context (C)}: True if C is a context within the assurance case. This predicate can be represented as Context (ID, Description) where ID is the unique identifier for the context and description is the textual description of the context.
             
             \item \textbf{Assumption (A)}: True if A is an assumption within the assurance case. This predicate can be represented as Assumption (ID, Description) where ID is the unique identifier for the assumption and description is the textual description of the assumption.
             
             \item \textbf{Justification (J)}: True if J is a justification within the assurance case. This predicate can be represented as Assumption (ID, Description) where ID is the unique identifier for the assumption and Description is the textual description of the assumption.
             
             \item \textbf{Undeveloped (X)}: True if X is either a Goal or Strategy marked as undeveloped. This predicate is represented as Undeveloped(X), where X can be either a goal or strategy.
           \newline
             
        \end{itemize}

         \subsubsection{Predicates for Formalizing an Assurance Case Pattern} 
         To formalize an assurance case pattern complying with  GSN, we propose the following predicates:
        \begin{itemize}
            \item \textbf{Uninstantiated (X)}: True if element X (can be any GSN element) is marked as uninstantiated.
            \item \textbf {UndevelopStantiated(X)}: True if element X is either a Goal or Strategy and is marked both as uninstantiated and undeveloped. 
            
            \item \textbf{HasPlaceholder (X)}: True if element X (can be any GSN element) contains a placeholder ‘\{\}’ within its description that needs instantiation. 
            
            \item \textbf{HasChoice (X, [Y], Label)}: True if an element X (either a Goal or Strategy) can be supported by selecting among any number of elements in [Y] (where Y can be any GSN element) according to the cardinality specified by an optional Label. The label specifies the cardinality of the relationship between X and Y. A label is of the general form \textit{m of n} (e.g. a label given as \textit{1 of 3} implies an element in X can be supported by any one of three possible supporting elements in [Y]).

            \item \textbf{HasMultiplicity (X, [Y], Label)}: True if multiple instances of an element X (either a Goal or Strategy) relate to multiple instances of another element [Y] (where Y can be any GSN element) according to the cardinality specified by an optional Label. The label specifies 
            how many instances of an element in X relate with how many instances of an element in [Y] (e.g.\textit{ m of n} implies \textit{m} instances of an element in X must be supported by \textit{n} instances of an element in Y).

            \item \textbf{IsOptional (X, [Y], Label)}: True if an element X (either a Goal or Strategy) can be optionally supported by another element [Y] (where Y can be any GSN element) according to the cardinality specified by an optional Label. The label specifies the cardinality of the relationship between X and Y (i.e. an instance of an element in X may be supported by another instance of an element in [Y], but it is not required). \newline
            
        \end{itemize}
        \subsubsection{ Predicates for Formalizing Relationships between GSN elements}
        
        The predicates we propose below are analogous to the two core GSN relationships i.e.  \textit{InContextOf} and \textit{SupportedBy}. 
        
        \begin{itemize}
            \item \textbf{IncontextOf (X, [N], D)}: True if element X at depth D has a neighbor [N] to the left or right at depth D, where N can be an assumption, justification, or context. X can be a goal or strategy and D represents the height or depth of the goal or strategy element and its neighbors in the GSN hierarchical structure. 

            \item \textbf{SupportedBy (X, [C], D)}: True if element X at depth D has children [C] directly below it, where [C] can include Goal(G), Strategy(S), or Evidence(E) and X can be a goal(G), or strategy(S). 
            \begin{itemize}
                \item If X is a Strategy, [C] can only be a Goal. 
                \item If X is a Goal, [C] can be either Goal, Strategy, or Evidence.
            \end{itemize}

        \end{itemize}


\subsection{Phase II: Data Collection}
\label{data-collection-subsection}
In our previous work \citep{b125}, 
we conducted a bibliometric analysis on assurance case patterns. This allowed us to collect 92 primary studies published within the past two decades and focusing on assurance case patterns. To collect data relevant for our current study, we therefore analyzed these 92 studies, then  selected five of them provided they described a pattern(s) and assurance cases instantiated (derived) from that pattern(s). We also made sure the selected patterns and assurance cases covered various application domains. This allowed us to create a dataset consisting of a set of assurance case patterns together with assurance cases derived from them. 


\subsection{Phase III: Data Pre-processing}

\begin{figure} [!h]
    \centering
    \includegraphics[width=0.7\linewidth]{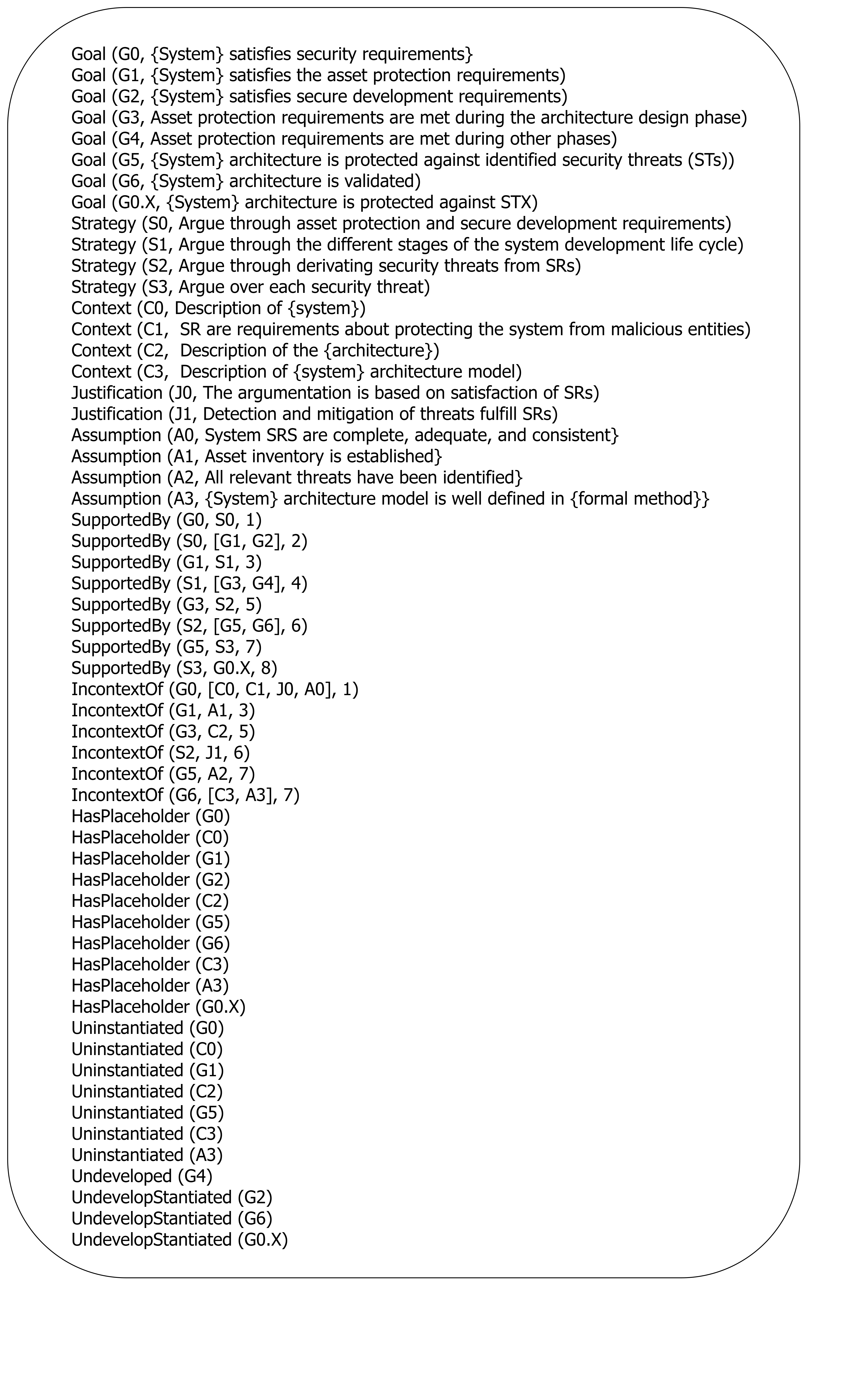}
    \caption{A Simple Predicate-based Representation of the Assurance Case Pattern depicted in Figure \ref{ACAS_XU_ACP} (see appendix).}
    \label{Formalized_ACP}
\end{figure}

Based on the defined predicates for GSN elements, relationships, and decorators we proposed in Phase I, we convert each collected assurance case pattern in GSN to a corresponding predicate form that an LLM can understand. 
This conversion facilitates the LLMs' comprehension of the inherent tree-like structure and relationships among the GSN elements in our dataset. It also aids in the generation of assurance cases complying with specific assurance case pattern(s). Figure \ref{Formalized_ACP} illustrates a predicate-based representation of an assurance case pattern in our dataset. 

\subsection{Phase IV: Using LLM to Automatically Generate Assurance Cases}
To generate assurance cases from patterns, we rely on LLMs. Each LLM takes as input the predicate-based representations of assurance case patterns and uses this representation as rules to: 1) enhance its reasoning capabilities by learning the features of patterns that are typically used to manually generate assurance cases and; 2) guide the automatic instantiation of assurance cases from the formalized patterns specified as inputs to the LLM. Each LLM generates an assurance case in the traditional structured prose (not in the predicate-based format). This allows using GSN guidelines \citep{b22} to turn the generated assurance case into GSN diagrams. 

To support the generation process, we rely on prompt engineering to provide instructions, guidelines, and enforce rules to ensure a desired generated response. We describe the LLMs prompts in Section \ref{sec:Prompt_Desc}.

\section{Experimental Setup}
\label{sec:Setup}
\subsection{Research Questions}
\label{sec:research-questions}

Our goal is to explore the potential of generative AI, particularly LLMs, in facilitating the automatic generation of assurance cases complying with specific assurance case patterns. To achieve this, we investigate  three research questions (RQs): 

\textbf{(RQ1): Are LLMs capable of creating well-formed and semantically valid assurance cases when they do not have SE (software engineering) knowledge specified in the prompts?} In this RQ, we assess the effectiveness of LLMs in creating assurance cases for a given system without providing SE knowledge to the LLMs. 

\textbf{(RQ2): Are LLMs capable of automatically instantiating assurance cases from assurance case patterns when SE knowledge is specified in the prompts?} In this RQ, we investigate the impact of providing various types of SE knowledge to our LLMs through prompt engineering. We assess the performance of these LLMs in generating assurance cases based on a given assurance case pattern. 
The categories of software engineering (SE) knowledge are: 1) examples; 2) domain information; 3) contextual information; 4) predicate-based rules; 5) a combination of the aforementioned four categories of knowledge. Thus, to facilitate the reasoning about this research question, we further break it into four sub-research questions:
\newline\indent\indent \textbf{(RQ2.1): Which impact does an example have on the performance of LLMs when it comes to generating assurance cases?}
\newline\indent\indent \textbf{(RQ2.2): Which impact does the domain information have on the performance of LLMs when it comes to generating assurance cases?}
\newline\indent\indent \textbf{(RQ2.3): Which impact does contextual information have on the performance of LLMs when it comes to generating assurance cases?}
\newline\indent\indent \textbf{(RQ2.4): Which impact do predicate-based rules have on the performance of LLMs when it comes to generating assurance cases?}

\textbf{(RQ3): Which of the evaluated LLMs performs best when it comes to automatically instantiating assurance cases from assurance case patterns?} In this RQ, we compare the performance of the analyzed LLMs (i.e. GPT-4 Turbo, and GPT-4o) in the automatic instantiation of assurance case from patterns.

\subsection{Description of the dataset used in the experiments}
\label{sec:dataset}

Our dataset comprises six assurance case patterns and five corresponding assurance cases that comply with these patterns, covering five distinct systems. These systems span various application domains, namely: the aviation, automotive, medical, and computing domains. We selected one assurance case pattern, presented in our formalized format, along with the assurance case complying with this pattern, presented in a structured prose format, as a one-shot example for our experiments. This example helps to illustrate to our LLM the concept of generating assurance cases from assurance case patterns. The remaining assurance case patterns, presented in formalized format, are used in the LLM prompts. This helps LLMs to generate assurance cases that comply with these patterns. Table \ref{table:dataset} provides various statistics,  such as the count of decorators (e.g., \textit{undeveloped}, \textit{uninstantiated}, \textit{choice}) in the assurance case patterns, and the count of relationships (e.g., \textit{InContextOf}, \textit{SupportedBy}) in the corresponding assurance cases that comply with these patterns. In the remainder of this section. we provide a detailed description of our dataset. For each system, we explain the assurance case and associated pattern(s) used to manually develop the assurance case. These manually created assurance cases serve as our ground-truth data for evaluating the LLM generated assurance cases.

The GSN diagrams depicting the ACPs and ACs complying with these patterns for the systems in our dataset are available in our replication package \footnote{\url{https://doi.org/10.6084/m9.figshare.27103225.v2}}.

\begin{table} [!h]
\centering
\caption {Overview of our dataset}
\label{table:dataset}

 \begin{tabular}{|p{2cm} |p{2.0cm} |p{2cm} |p{2.0cm}  |p{1.5cm}  |p{1.5cm}  |p{1.8cm}|}

\hline
\multirow{2}{*}{System} & \multirow{2}{*}{Domain}& \multicolumn{3}{c|}{Assurance Case Patterns (ACPs)} & \multicolumn{2}{c|}{Assurance Cases (ACs)}\\
 
\cline{3-7}
 & & \textbf{Decorators}& \textbf{Placeholders}& \textbf{Elements}& \textbf{Elements}& \textbf{Relationships}\\
\hline
ACAS XU & Aviation & 11 & 10 & 22 & 24 & 23 \\
\hline
BLUEROV2 & Automotive & 17 & 8 & 18 & 24 & 21 \\
\hline
GPCA & Medical & 6 & 21 & 23 & 27 & 26 \\
\hline
IM SOFTWARE & Computing & 1 & 9 & 15 & 24 & 23 \\
\hline
DEEPMIND & Medical & 16 & 26 & 17 & 23 & 23 \\
\hline

\end{tabular}

\end{table}

\subsubsection{ACAS XU and Its Assurance Framework}

ACAS Xu (Airborne Collision Avoidance System Xu) is a collision avoidance system designed for use in unmanned aerial vehicles (UAVs), commonly known as drones \citep{b126}. 
To ensure that ACAS Xu is acceptably secure against security threats, \citet{b126} provided a threat identification assurance case pattern. They demonstrated the application of this pattern in creating a partial security case specific to the ACAS Xu system. 


\subsubsection{BlueROV2 and Its Assurance Framework}

The BlueROV2 system is an advanced Unmanned Underwater Vehicle (UUV) or underwater Remotely Operated Vehicle (ROV) \citep{b88}. Its main objective is to autonomously track pipelines on the seafloor while avoiding static obstacles such as plants and rocks \citep{b88}. Safety assurance for  BlueROV2 is achieved through the identification of potential hazards and reduction of the risk posed by those hazards based on the ALARP (As Low As Reasonably Practicable) principle \citep{b88}. \citet{b88}, generated an assurance case for BlueROV2 using the ALARP pattern sequentially composed with another pattern called the ReSonAte pattern. We utilized these two patterns and the generated assurance case in our dataset. 
Figure \ref{BlueROV2_ACP} (located in the appendix)  shows the combined pattern formed from both the ALARP pattern and the ReSonAte pattern. The \textit{SupportedBy} relationship between G3 and S4 in that Figure links the two patterns together. 

\subsubsection{GPCA and Its Assurance Framework}
The Generic Patient-Controlled Analgesia (GPCA) system, also known as an \textit{Infusion pump} is one of the most common safety critical systems in the medical domain. Some of the operational hazards faced by this system include ``\textit{Overinfusion}'' and ``\textit{Underinfusion}'' which can have dire consequences for patient safety \citep{b9}. To demonstrate the application of assurance cases for certifying the safety of critical systems, several studies \citep{b101, b9} have utilized the GPCA system as a case study. 

\citet{b9}  presented a safety case pattern and a safety case for the GPCA system complying with this pattern. 
Note that some GSN elements of this safety case have duplicated identifiers whereas GSN usually fosters uniqueness of identifiers.
 To mitigate that duplication, we systematically reassigned unique identifiers to each duplicated GSN element.

\subsubsection{The Instant Messaging (IM) Server Software and Its Assurance Framework}

The Instant messaging (IM) server software is  used for information exchange, with the data within the software forming the basis of user interaction \citep{b11}. That system is characterized by its independent behavioral features, known as its internal structure, as well as by its interactive relationships with external components, referred to as the external manifestation (EM) \citep{b11}. The EM encompasses the overall interaction of the software with the outside world, including the set of external environment entities, the interaction set between these entities and the software, and the direction of these interactions.
The internal structure (IS) of software focuses on the internal functional processes and data transmission within the software. This includes the set of software functional processes, data storage, and internal interaction sets \citep{b11}.

To ensure that the IM server software is acceptably secure requires a demonstration to show that critical assets such as user account information, and authentication information are well-protected \citep{b11}. Thus, \citet{b11} presented a software security top-level argument pattern and utilized this pattern to create a software security case for the IM software. 
Some GSN elements of this security case are duplicated. To mitigate the duplication as explained above, we systematically renumbered and reassigned unique identifiers to each GSN element.

\subsubsection{The DeepMind ML system for retinal disease diagnosis and Its Assurance Framework}
\label{Section:deepmind}
The DeepMind system is an example of a safety-critical system that uses Machine Learning based functionality. The DeepMind system utilizes two neural networks to predict retinal disease from eye scans. The first neural network processes a retinal scan to generate a tissue-segmentation map. This map is then analyzed by the second neural network, which provides a diagnosis and referral \citep{b108}.

\citet{b108}  presented an assurance case pattern for justifying the sufficiency of the interpretability of ML in safety-critical systems. They demonstrated the application of this pattern in creating an assurance case for the interpretability of the machine learning component in the DeepMind system. 

We utilized both this ACP and AC generated for the ML component of the DeepMind system as a one-shot example in our experiments. Our choice of that example is random. 

\subsection{Large Language Models Setups}
\label{section-llm-setups}
To carry out our experiments, we focused on two LLMs, namely: GPT-4o and GPT-4 Turbo. 
We chose them because they are powerful and incorporate the latest features. The non-deterministic nature of LLMs, motivated us to run each of our experiments multiple times i.e. \textit{K} times, where \textit{K = 5}. This allows mitigating the potential inconsistencies in responses and ensuring reliable evaluation of our LLMs. To interact with both LLMs, we relied on the OpenAI API \citep{b112}. 
 We set the default values for the following parameters when interacting with both LLMs:
\begin{itemize}

    \item \textbf{The temperature}: it controls the randomness and creativity in the output of LLMs. By adjusting this parameter, users can balance creativity and coherence in the generated text \cite{b113}. 
    In our experiments, we set the temperature parameter to its default value of 1. 
 
    \item \textbf{The maximum number of tokens}: It controls the length of responses generated by the LLM. In our experiments, we consistently set its value to ``4096'' to accommodate longer text across all experiments.
\end{itemize}

\subsection{Description of the Experiments and the supporting information}

In our experiments, we applied the Chain-of-Thought (CoT) \citep{b66}  prompting technique. This allows enhancing the reasoning capabilities of the LLMs and therefore improves their ability to perform a complex reasoning task, namely: the generation of assurance cases.

\subsubsection{Description of the supporting information}
\label{sec:Supp_Info}

As stated previously, we can specify several categories of SE knowledge in our prompts. Like in the literature (e.g., \citep{b75},  \citep{b107} and  \citep{b114}), to allow each LLM to perform effectively and generate outputs (assurance cases) close to the ground-truth, we notably rely on the following two categories of SE knowledge:

 \begin{itemize}
        \item \textbf{Contextual Information} – We define \textit{'contextual information'} as the background details conveying the fundamental information about the structure and representation of the different elements and decorators in the assurance case and assurance case pattern represented in GSN. The contextual information also provides instructions on how to derive an assurance case from an assurance case pattern. It aims at allowing the LLM to enhance its ability to interpret and understand the general structure, content, and guidelines necessary to generate assurance cases complying to a given pattern effectively. It  remains the same across all our experiments. 
        The contextual information we used in our experiments is available in the Appendix (see Section \ref{appendix:contextual-information}).

        \item \textbf{Domain Information} – In our experiments, \textit{'domain information'} refers to the specialized knowledge, terminology, and facts specific to the domain or system for which an assurance case is being automatically created. Examples of this information include details about the mode of operation, test results, and verification activities within that domain or system. The domain information enables the LLM to select from a variety of artifacts (arguments, evidence) necessary to replace the generic information found in placeholders within the assurance case pattern. The domain information can vary from one application domain to another. Hence, in our experiments, we utilized different domain information for each system in our dataset. We extracted this domain information from the cited references \citep{b88, b9, b11, b126} that describe our dataset. 
        
    \end{itemize}

\subsubsection{Description of the Experiments}

Table \ref{tab:experiments} provides a descriptive summary of the various experiments in our study. Each row, labeled from Experiment 1 to Experiment 9, describes the individual configuration of each experiment, while each column represents one of four distinct categories of software engineering (SE) knowledge: Example, Context Information, Domain Information, and Predicate Rules. The presence of each of these categories in a given experiment is denoted by an 'X' mark in the corresponding cell. For instance, Experiment 1 excludes all four categories, Experiment 2 includes all four, while Experiment 9 only includes Predicate Rules.

\paragraph{Experiment without software engineering knowledge specified in the prompts}
In this experiment (i.e. Experiment 1), to answer RQ1, we want to assess the performance of GPT-4 Turbo and GPT-4o in generating assurance cases when no extra software engineering knowledge is included in their prompts. Hence, in this experiment, we do not provide context information, domain information, examples, and predicate rules to both LLMs. 

\paragraph{Experiments with SE knowledge specified in the prompts}
Building on previous work from literature (e.g., \citep{b114}, \citep{b75}, and \citep{b107} ) and to answer our RQ2, we conducted eight additional experiments (i.e. Experiments 2 to Experiments 9), each leveraging at least one category of SE knowledge. 

\begin{table}
    \centering
        \caption{Overview of our Experiments}
    \label{tab:experiments}
    \begin{tabular}{ccccc}
         &  Example&  Context Information&  Domain 
         Information& Predicate Rules\\
         \hline
        \setcounter{enumi}{1}
        \refstepcounter{experiment}
        \label{exp:1}Experiment 1&  &  &  & \\
         \hline
         \hline
         \refstepcounter{experiment}\label{exp:2}Experiment 2&  x&  x&  x& x\\
         \hline
        \refstepcounter{experiment}\label{exp:3}Experiment 3&  &  x&  x& x\\
         \hline
         \refstepcounter{experiment}\label{exp:4}Experiment 4&  x&  x&  & x\\
         \hline
         \refstepcounter{experiment}\label{exp:5}Experiment 5&  &  x&  & x\\
         \hline
        \refstepcounter{experiment}\label{exp:6}Experiment 6&  x&  &  x& x\\
         \hline
         \refstepcounter{experiment}\label{exp:7}Experiment 7&  &  &  x& x\\
         \hline
         \refstepcounter{experiment}\label{exp:8}Experiment 8&  x&  &  & x\\
         \hline
          \refstepcounter{experiment}\label{exp:9}Experiment 9&  &  &  & x\\
         \hline
    \end{tabular}

\end{table}

Note that all the patterns we use as input are represented in the predicate-based format that we specified. To allow both LLMs to digest that format, we therefore specify predicate rules in each of the eight experiments described above.

\subsection{Description of the structure of the prompts used in the experiments}
\label{sec:Prompt_Desc}

The OpenAI API supports three types of prompts: the system prompt, the user prompt, and the assistant prompt. These can be categorized into two main groups that we further explain below.

\subsubsection{Input passed to the LLM} 
The input passed to the LLM is the LLM prompt. The latter results from the combination of two other prompts: \begin{itemize}
        \item \textbf{System Prompt}:  This consists of instructions and guidelines provided to the LLM to ensure it responds appropriately. Depending on the experiment, our system prompt may contain all or a combination of the various categories of SE knowledge. Figure \ref{System_Prompt_No_SE} depicts the system prompt given to each LLM for Experiment \ref{exp:1}. That system prompt deliberately excludes any SE knowledge. This allows us to evaluate the inherent performance of the LLMs without the influence of SE knowledge.

        \begin{center}
\noindent\begin{tcolorbox}[colframe=black, colback=white, coltitle=white, arc=10pt, width=15cm, halign title=center]
\smallskip
\begin{center}

You are an assistant who assists in developing an assurance case in a tree structure using Goal Structuring Notation (GSN). Your role is to create an assurance case.

\end{center}
\end{tcolorbox}
 \captionof{figure}{A Sample System Prompt for Experiment \ref{exp:1}}
 \label{System_Prompt_No_SE}
\end{center}

        \item \textbf{User Prompt}: This is the input or query from the user interacting with the model, requesting the model to complete a specific task. In the user prompts specified for experiment without SE knowledge (i.e. Experiment \ref{exp:1}),  We do not specify any assurance case pattern, as we consider this to be a form of SE knowledge. Figure \ref{User_Prompt_No_SE} shows a sample of our user prompt for Experiment \ref{exp:1}. 

        \begin{center}
\noindent\begin{tcolorbox}[colframe=black, colback=white, coltitle=white, arc=10pt, width=15cm, halign title=center]
\smallskip
\begin{center}

Create a security case for ACAS Xu (Airborne Collision Avoidance System Xu) and display it in a hierarchical tree format using dashes (-) to denote different levels.

\end{center}
\end{tcolorbox}
 \captionof{figure}{A Sample User Prompt for Experiment \ref{exp:1}}
 \label{User_Prompt_No_SE}
\end{center}
        
 \end{itemize}       
 
  \subsubsection{Output generated by the LLM}
    
 The \textbf{Assistant Prompt} is the output that the LLM generates. More specifically, the  assistant prompt refers to the response generated by the model based on the system and user prompts provided to the model. Thus, in our work, the assistant prompt provides the assurance case that the LLM generates. 

  \begin{figure} [!htb]
    \centering
    \includegraphics[width=0.9\linewidth]{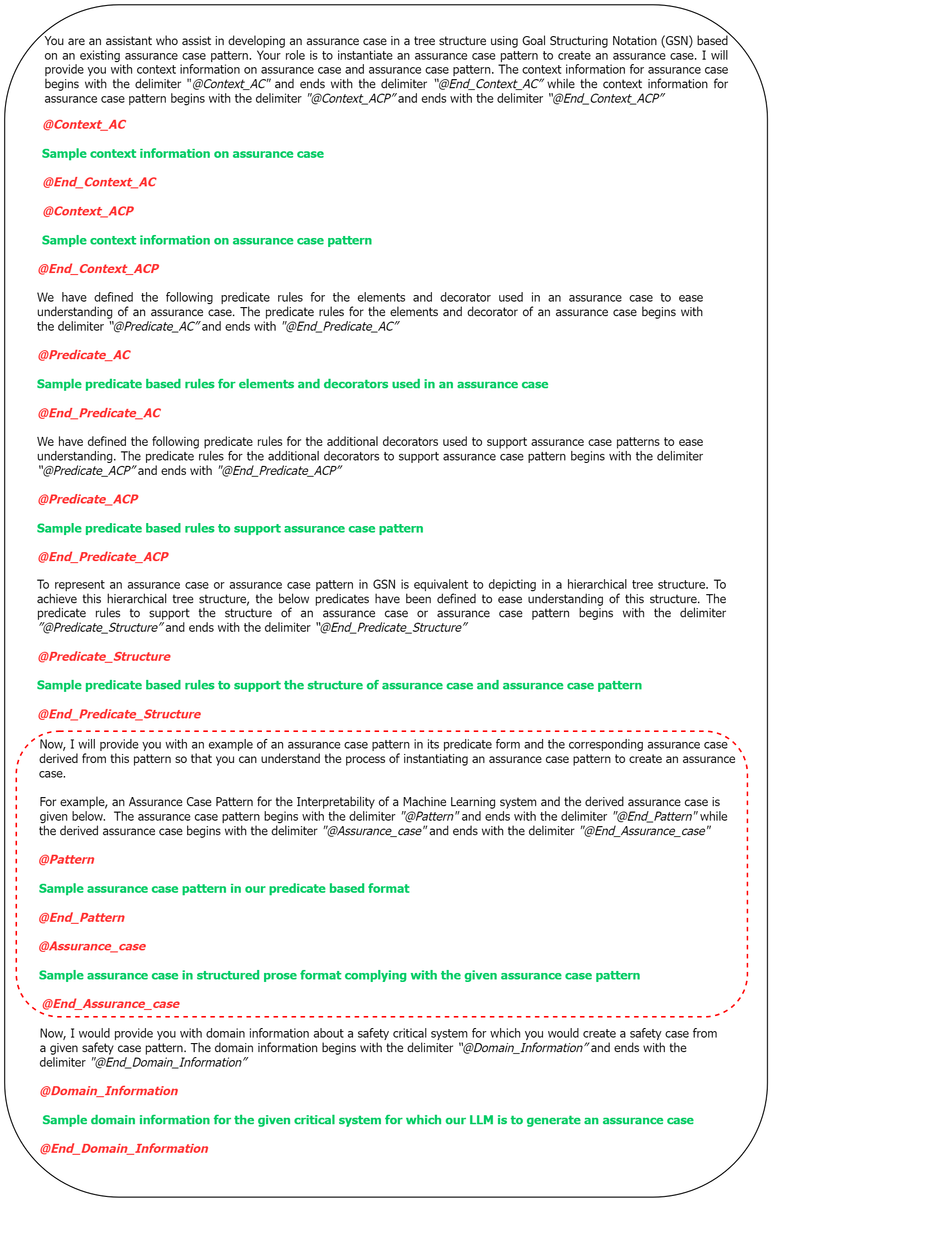}
    \caption{Generic structure of our System Prompts for Experiments with SE Knowledge}
    \label{SE_System_Prompt}
\end{figure}

\subsubsection{Description of System Prompts with S.E Knowledge}

Figure \ref{SE_System_Prompt} depicts the template (i.e. the generic structure) of the system prompts we used in our experiments with SE knowledge.

\paragraph {Description of Zero-shot System Prompts with SE Knowledge}
In our zero-shot experiments involving SE knowledge (i.e. Experiments \ref{exp:3}, \ref{exp:5}, \ref{exp:7}, and \ref{exp:9}), the system prompts used to query both LLMs may consist various categories of SE knowledge. However, these prompts exclude the example category, which is indicated by red dotted lines in Figure \ref{SE_System_Prompt}.

\paragraph {Description of One-shot System Prompts with SE Knowledge}
In our One-shot experiments involving SE knowledge (i.e. Experiments \ref{exp:2}, \ref{exp:4}, \ref{exp:6}, and \ref{exp:8}), the system prompts used to query both LLMs may consist in various categories of SE knowledge. This includes the example category, which is indicated by red dotted lines in Figure \ref{SE_System_Prompt}.

\subsubsection{Description of User Prompts with S.E Knowledge}

\begin{figure} [!htb]
    \centering
    \includegraphics[width=0.6\linewidth]{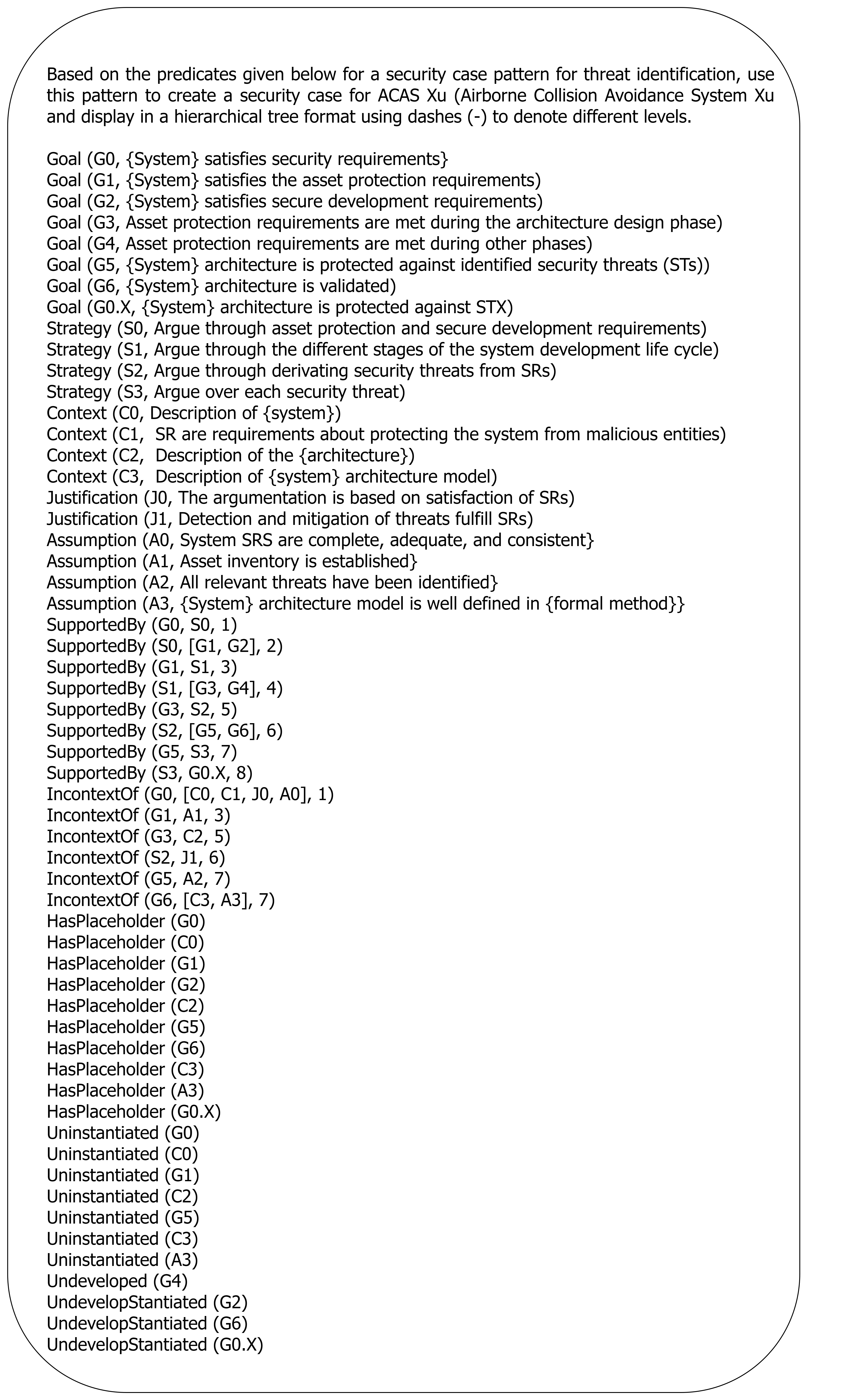}
    \caption{A Sample User Prompt for Experiments with SE Knowledge.}
    \label{SE_User_Prompt}
\end{figure}

In the user prompts specified for experiments involving SE knowledge (i.e., Experiments \ref{exp:2} to Experiment \ref{exp:9}), we include the assurance case pattern in our predicate-based format for each system. We then prompt the model to create an assurance case that complies with this pattern for a given system in our test dataset. Figure \ref{SE_User_Prompt} shows an excerpt of a sample user prompt in our experiments with SE knowledge.

\subsection{Evaluation Metrics}
Each evaluation metric we outline below allows  assessing the  similarity between the  experiments results (i.e. LLM-generated assurance cases) and the ground-truth.

\subsubsection{Exact Match}

As in \citet{b127}, we employ this metric to gauge the accuracy with which our LLMs generated output matches the ground truth, character by character, without discrepancies. The scoring typically ranges from 0 to 1, where 0 signifies no similarity and 1 indicates a perfect or near-perfect match. To compute this measure, we rely on the Python library called \textit{FuzzyWuzzy} \citep{b115}.

\subsubsection{BLEU Score}

As in \citet{b129}, to assess the similarity between the generated text and the ground-truth, we utilize the BLEU score. The latter is one of the widely used metrics in NLP (Natural Language Processing) and one of the most commonly used evaluation metrics for natural language texts \citep{b78}. It ranges between 0 and 1, where 0 indicates no match between the generated text and the ground-truth text, while 1 indicates a perfect match between both the generated text and the ground-truth text. We use a Python library called \textit{Sacrebleu} \citep{b116} for its assessment.

\subsubsection{Semantic Similarity}

In this metric, we evaluate how closely the texts in the GSN elements of the assurance cases generated by our LLMs relate in meaning to the ground-truth assurance cases. To assess that measure, we rely on the cosine similarity measure \citep{b118, b119}. 
 Cosine similarity values range from -1 to 1, where -1 indicates no similarity or completely dissimilar texts, and 1 indicates identical texts. 
We rely on a Python library called \textit{scikit-learn} \citep {b117} to automatically compute the values of that metric.

\section{Results}

\label{sec:Result}

\subsection{RQ1: Are LLMs capable of creating well-formed and semantically valid assurance cases when they do not have SE knowledge specified in their prompts?}
\subsubsection{Metric results}
Table \ref{tab1} reports the median of the Exact match, BLEU scores, and Cosine similarity results we obtained when running each LLM five times in Experiment \ref{exp:1}. The standard deviations of these results are very close to zero.  This indicates our results are stable across multiple runs. For brevity's sake, we do not report these deviations. 

The metric values Table \ref{tab1} reports are extremely low and therefore mediocre. More specifically, the exact match values in Table \ref{tab1} are notably very low, with the highest median exact match value being 0.05. On the other hand, both the BlueROV2 system under both models and IM Software system under GPT-4o yields the lowest median exact match value of 0.02. When it comes to the BLEU scores Table \ref{tab1} reports, we observe that GPT-4o yields the highest median BLEU score value of 0.03 across all runs for the ACAS XU system.  Finally, as Table \ref{tab1} shows, the semantic similarity results are slightly better compared to both the Exact match and BLEU score values, but still remain below average (i.e. 0.5). Note that the ACAS XU system under GPT-4o and DeepMind system under GPT-4o yields the highest semantic similarity median value of 0.23 across all runs in Experiment \ref{exp:1}.

\begin{table} [!htb]
\centering
\caption{Median Exact Match, BLEU Score, and Semantic Similarity Results for Experiments without SE Knowledge}
\label{tab1}
\begin{tabular}{|l |l |l |l |l|} \hline       
\textbf{System} & \textbf{Model} & \textbf{Exact Match} & \textbf{BLEU Score} & \textbf{Semantic Similarity} \\ \hline 
\multirow{2}{*}{ACAS XU} & GPT-4o & 0.04 & 0.03 & 0.23 \\ \cline{2-5}
 & GPT-4 Turbo & 0.03 & 0.02 & 0.22 \\  \hline
\multirow{2}{*}{BlueROV2} & GPT-4o & 0.02 & 0 & 0.08 \\ \cline{2-5}
 & GPT-4 Turbo & 0.02 & 0.01 & 0.07 \\  \hline
\multirow{2}{*}{GPCA} & GPT-4o & 0.04 & 0 & 0.04 \\ \cline{2-5}
 & GPT-4 Turbo & 0.04 & 0.01 & 0.11 \\  \hline
\multirow{2}{*}{IM Software} & GPT-4o & 0.02 & 0.01 & 0.09 \\ \cline{2-5}
 & GPT-4 Turbo & 0.05 & 0.01 & 0.13 \\  \hline
\multirow{2}{*}{DeepMind} & GPT-4o & 0.05 & 0.01 & 0.23 \\ \cline{2-5}
 & GPT-4 Turbo & 0.05 & 0.01 & 0.22 \\  \hline

\end{tabular}

\end{table}

\subsubsection{Reasons explaining Experiment \ref{exp:1} results}
The manual analysis of  the LLM-generated assurance cases obtained with Experiment  \ref{exp:1} allowed us to identify some reasons explaining the poor results it yields:

\begin{itemize}
    \item {\textbf{Lack of a clear goal structure}}: In Experiment \ref{exp:1}, the majority of LLM-generated assurance cases lack the goal-structured hierarchy required for clarity. These assurance cases list GSN elements in an unclear and unstructured manner, failing to capture the relationships between them. This random organization of elements deviates from the structural rules defined in the GSN standard \cite{b22}, causing confusion about the statements in goals, the strategies for achieving these goals, and the supporting evidence.

    \item {\textbf{Inconsistent GSN Element names and IDs:}} In some of the LLM-generated assurance cases Experiment \ref{exp:1} yields, there are inconsistencies in GSN element names and the symbols the LLMs use to identify them. For instance, when generating the assurance case for ``ACAS XU'' in Run 1, GPT-4 Turbo used the same symbol \textit{``S''}  to represent both a strategy and a solution, whereas assurance case developers usually use two different symbols/abbreviations to name and distinguish strategies from solutions. Likewise, when generating the assurance case for "DeepMind" in Run 2, GPT-4o used a variety of inconsistent abbreviations ( i.e. \textit{``su'', ``sp'', ``st'', ``sd'', ``sn'', ``sm'', ``sl''}) to represent solutions, leading to ambiguity in the assurance cases. Assurance case developers usually use a unique symbol/abbreviation to name solutions.

    Furthermore, when generating the assurance case of  BlueROV2 for Run 2, the assurance case that GPT-4o generated included an element called \textit{``Evidence''} in addition to a \textit{``Solution''} element. This can lead to confusion as \textit{``Evidence''} is not a GSN concept. Similarly, when generating the assurance case of BlueROV2 system for Run 2, GPT-4 Turbo  termed an element  \textit{``Argument''}, whereas \textit{``Argument''} is not a GSN concept. Likewise, when generating the assurance case for the GPCA system  for Run 5, GPT-4 Turbo generated an assurance case that includes an unexpected element called \textit{``Inference''}. Both \textit{``Argument''} and \textit{``Inference''} are not recognized as GSN elements (i.e. concepts) within the GSN standard. Thus, this further contributes to inconsistency and potential misinterpretation of the elements within an assurance case.

    \item {\textbf{Absence of GSN Element Identifier:}} In some of the LLM-generated assurance cases, there are no element IDs. For instance, when generating the assurance case of  ``BlueROV2'' for Run 2,  GPT-4o  included in that assurance case a list of elements  having no IDs.  This absence of unique identifiers can impede the understanding of inferential links between GSN elements. It can also make challenging the understanding of how the evidence (through solutions) supports the arguments within an assurance case. 
\end{itemize}

\smallskip
\noindent\fbox{%
    \parbox{\linewidth}{%
    \smallskip
In summary, our analysis of Experiment \ref{exp:1} for \textbf{RQ1} reveals that without SE knowledge, LLMs cannot generate reliable assurance cases. The exact match values are very low, with the highest median exact match value being only 0.05. The highest median BLEU scores is 0.03, and the best semantic similarity value is 0.23. These values indicate that the generated assurance cases significantly differ from the ground-truth assurance cases, rendering them ineffective for supporting system assurance. This emphasizes the need for incorporating SE knowledge to improve the reliability of LLM-generated assurance cases.
    }}

\subsection{RQ2: Are LLMs capable of automatically instantiating assurance cases from assurance case patterns with SE knowledge specified in the prompts?}
\label{sec:RQ2_Result}

\subsubsection{Metric results}
Similar to the methodology used by \citep{b75, b130}, we performed each of our experiments five times (K = 5). Therefore, across the 8 experiments involving SE knowledge, both GPT-4o and GPT-4 Turbo generated a total of 320 assurance cases for our test systems.
Tables \ref{tab2}, \ref{tab3}, and \ref{tab4} respectively report the median of the Exact match results, BLEU scores and Semantic similarity results we obtained across our 8 experiments involving SE knowledge (i.e., Experiment \ref{exp:2} to Experiment \ref{exp:9}), using the AC and ACP of DeepMind as the one-shot example. Each row in these tables corresponds to a system in our test dataset under both GPT-4o and GPT-4 Turbo. In the eight experiments, the  standard deviations associated with metrics values are close to zero, indicating a stability in these values. Thus, for brevity's sake, we do not report them. We discuss the results reported in these three tables in the remainder of this section. 


\begin{table} [!htb]
\centering
\caption{Median Exact Match Result for Experiments with Software Engineering Knowledge}
\label{tab2}
\begin{tabular}{|l |l |l |l |l |l |l |l |l |l|} \hline   
\textbf{System} & \textbf{Model} & \textbf{E2} & \textbf{E3} & \textbf{E4} & \textbf{E5} & \textbf{E6} & \textbf{E7} & \textbf{E8} & \textbf{E9} \\ \hline 
\multirow{2}{*}{ACAS XU} & GPT-4o & 0.79 & \textbf{0.85} & 0.84 & 0.82 & 0.77 & 0.74 & 0.83 & 0.82 \\  \cline{2-10}
 & GPT-4 Turbo & 0.65 & 0.59 & \textbf{0.83} & 0.81 & 0.56 & 0.65 & 0.52 & 0.69 \\ \hline
\multirow{2}{*}{BlueROV2} & GPT-4o & 0.65 & 0.78 & 0.75 & 0.75 & \textbf{0.8} & 0.75 & 0.77 & 0.71 \\  \cline{2-10}
 & GPT-4 Turbo & 0.76 & 0.76 & 0.66 & 0.58 & \textbf{0.81} & 0.78 & 0.64 & 0.61 \\ \hline
\multirow{2}{*}{GPCA} & GPT-4o & 0.26 & 0.3 & 0.26 & 0.34 & 0.22 & 0.28 & \textbf{0.35} & \textbf{0.35} \\  \cline{2-10}
 & GPT-4 Turbo & 0.25 & 0.31 & 0.3 & 0.21 & 0.23 & 0.32 & \textbf{0.32} & 0.25 \\ \hline
\multirow{2}{*}{IM Software} & GPT-4o & 0.1 & 0.1 & 0.12 & 0.12 & 0.1 & \textbf{0.28} & 0.12 & 0.27 \\  \cline{2-10}
 & GPT-4 Turbo & 0.09 & 0.1 & 0.07 & 0.09 & 0.08 & \textbf{0.14} & 0.08 & 0.08 \\ \hline

\end{tabular}

\end{table}

\begin{table} [!htb]
\centering
\caption{Median BLEU Score Result for Experiments with Software Engineering Knowledge}
\label{tab3}
\begin{tabular}{|l |l |l |l |l |l |l |l |l |l|} \hline  
\textbf{System} & \textbf{Model} & \textbf{E2} & \textbf{E3} & \textbf{E4} & \textbf{E5} & \textbf{E6} & \textbf{E7} & \textbf{E8} & \textbf{E9} \\ \hline 
\multirow{2}{*}{ACAS XU} & GPT-4o & 0.73 & 0.75 & 0.68 & 0.68 & \textbf{0.76} & 0.72 & 0.68 & 0.64 \\  \cline{2-10}
 & GPT-4 Turbo & \textbf{0.71} & 0.4 & 0.69 & 0.6 & 0.7 & 0.41 & 0.67 & 0.65 \\ \hline
\multirow{2}{*}{BlueROV2} & GPT-4o & 0.54 & 0.53 & 0.55 & 0.52 & \textbf{0.57} & 0.56 & 0.55 & 0.53 \\  \cline{2-10}
 & GPT-4 Turbo & 0.58 & 0.46 & 0.5 & 0.4 & \textbf{0.62} & 0.56 & 0.55 & 0.51 \\ \hline
\multirow{2}{*}{GPCA} & GPT-4o & \textbf{0.32} & 0.28 & 0.27 & 0.23 & 0.21 & 0.31 & 0.26 & 0.23 \\  \cline{2-10}
 & GPT-4 Turbo & 0.16 & 0.22 & 0.19 & 0.17 &\textbf{ 0.25} & 0.2 & 0.22 & 0.15 \\ \hline
\multirow{2}{*}{IM Software} & GPT-4o & 0.27 & 0.27 & \textbf{0.3} & 0.29 & 0.27 & 0.29 & 0.22 & 0.23 \\  \cline{2-10}
 & GPT-4 Turbo & 0.17 & 0.19 & 0.18 & 0.16 & \textbf{0.21} & 0.18 & 0.16 & 0.17 \\ \hline

\end{tabular}

\end{table}

\begin{table} [!htb]
\centering

\caption{Median Semantic Similarity Result for Experiments with Software Engineering Knowledge}
\label{tab4}
\begin{tabular}{|l |l |l |l |l |l |l |l |l |l|} \hline    
\textbf{System} & \textbf{Model} & \textbf{E2} & \textbf{E3} & \textbf{E4} & \textbf{E5} & \textbf{E6} & \textbf{E7} & \textbf{E8} & \textbf{E9} \\ \hline 
\multirow{2}{*}{ACAS XU} & GPT-4o & \textbf{0.92} & \textbf{0.92} & 0.91 & 0.91 & \textbf{0.92 }& 0.9 & 0.91 & 0.85 \\ \cline{2-10}
 & GPT-4 Turbo & \textbf{0.91} & 0.89 & \textbf{0.91} & \textbf{0.91} & 0.9 & 0.85 & 0.87 & 0.9 \\ \hline 
\multirow{2}{*}{BlueROV2} & GPT-4o & 0.88 & 0.9 & 0.64 & 0.67 & \textbf{0.92} & 0.83 & 0.73 & 0.59 \\ \cline{2-10}
 & GPT-4 Turbo & 0.89 & 0.89 & 0.57 & 0.58 & \textbf{0.9} & 0.87 & 0.53 & 0.63 \\ \hline 
\multirow{2}{*}{GPCA} & GPT-4o & \textbf{0.81} & 0.76 & 0.37 & 0.28 & 0.73 & 0.76 & 0.37 & 0.27 \\ \cline{2-10}
 & GPT-4 Turbo & 0.56 &\textbf{ 0.57} & 0.28 & 0.27 & \textbf{0.57} & 0.55 & 0.31 & 0.27 \\ \hline 
\multirow{2}{*}{IM Software} & GPT-4o & 0.7 & 0.7 & 0.58 & 0.59 & \textbf{0.71} & \textbf{0.71} & 0.57 & 0.54 \\ \cline{2-10}
 & GPT-4 Turbo & 0.65 & \textbf{0.66} & 0.5 & 0.52 & 0.65 & 0.64 & 0.52 & 0.5 \\ \hline 

\end{tabular}

\end{table}

Table \ref{tab2} shows that the median Exact match results are quite high for both the ACAS XU and BlueROV2 systems. However, these results are quite low for the GPCA system, while the IM software system has the lowest Exact match values.



As shown in Table \ref{tab3}, the median of BLEU score results ranges from moderately high to moderate for both the ACAS XU and BlueROV2 systems. However, the scores are relatively low for both the GPCA system and IM software system.


 Table \ref{tab4} reports the semantic similarity results obtained for the systems in our test dataset. Thus, for each experiment, the semantic similarity results are outstanding for the ACAS XU system and high for the BlueROV2 system. This indicates the assurance cases the LLMs generated for both systems are semantically close to the ground-truth assurance cases. For both the GPCA and IM Software systems, the results vary between moderate and low depending on the type of experiment and the LLM utilized. It is worth noting that the semantic similarity results are relatively higher compared to the BLEU score and Exact match measure for the majority of our experiments. This is probably because the semantic similarity measure mainly considers the meaning of the texts associated with the elements comprised in the generated assurance cases. Thus, if two texts have similar meanings but different wording, they can still have a high cosine similarity. 


Potential reasons for the low results associated with GPCA and IM software are mainly ACP-related and may include the following:

\begin{itemize}
    \item \textbf{Cardinality Ambiguity in Multiplicity Relationship}: In an ACP, cardinality specifies the required number of instances of a particular element that must be associated with other elements within the pattern. However, the common use of generic labels such as (\textit{"0...*", "1....*", "N"}) to specify the cardinality of the multiplicity relationship in a pattern allows for various interpretations by LLMs. Hence, if the LLM at hand is not able to properly interpret that cardinality, this may lead to mismatches in the number of branches or relationships between elements of the generated assurance case compared to the ones in the ground-truth assurance cases. Additionally, due to the ambiguity in the cardinality of multiplicity relationships within patterns, we observed that LLMs occasionally generate duplicated GSN elements across the goal structure. For example, the same goal  may be generated multiple times with identical IDs and descriptions, or descriptions that are nearly identical. This may result in a discrepancy between the GSN elements generated by LLMs and the ground-truth GSN elements. 

    \item \textbf{Mismatch in Instantiating Abstract Parameters with Multiple Available Values}: In an ACP, a single abstract parameter can be instantiated with multiple concrete values, depending on the multiplicity within the pattern structure \citep{b88}. For elements with generic placeholders that can be replaced with diverse information from the available domain information, a mismatch in the number of branches or relationships between elements of the generated assurance case can occur. This mismatch is caused by the generic label for the cardinality of the multiplicity relationship and may result in a mismatch when instantiating multiple branches with multiple concrete values.
    \item \textbf{The complexity of the pattern:}  that complexity might affect the performance and efficiency of the LLMs. For example, the number of placeholders in the assurance case pattern for ACAS XU and BLUEROV2 systems are 10 and 8, respectively while the number of placeholders in the assurance case pattern for the GPCA system and IM software system are 21 and 9, respectively. This could impact the performance of LLMs in generating ACs close to the ground-truth ACs especially when there is a cardinality ambiguity in the multiplicity relationships in the input assurance case pattern.
    
\end{itemize}

Figures \ref{4o_BlueROV2_AC} and \ref{4Turbo_BlueROV2_AC} (see Appendix) both illustrate the effects of cardinality ambiguity in multiplicity relationship. Figure \ref{4Turbo_BlueROV2_AC} illustrates a graphical notation (i.e., a GSN representation) of the assurance case that GPT-4 Turbo generated for the BlueROV2 system. Figure \ref{4o_BlueROV2_AC} illustrates a graphical notation of the assurance case that GPT-4o generated for the same system and for the same experiment (i.e., Experiment \ref{exp:2}) \footnote{GPT-4 Turbo and GPT-4o generated both figures \ref{4o_BlueROV2_AC} and \ref{4Turbo_BlueROV2_AC} in structured prose format. Thus, in accordance with the GSN standard guidelines \citep{b22}, we have converted both Figures into a graphical notation (GSN).}. To facilitate the discussion and reasoning about these two figures, we compare both figures with the ground-truth assurance case of the BlueROV2 system that Figure \ref{BlueROV2_AC} depicts.  



The differences between the assurance cases generated by GPT-4o and GPT-4 Turbo for the BLUEROV2 system are significant. Figures \ref{4o_BlueROV2_AC} and \ref{4Turbo_BlueROV2_AC} highlight these differences: GPT-4o generated an assurance case with 42 GSN elements, whereas GPT-4 Turbo generated an assurance case with only 18 GSN elements. In contrast, the ground-truth assurance case, depicted in Figure \ref{BlueROV2_AC}, consists of 24 GSN elements.

This discrepancy in the number of GSN elements generated by GPT-4o and GPT-4 Turbo, as compared to the ground-truth assurance case, can be attributed to the ambiguity in the cardinality  (\textit{``1....*''}, \textit{``0....*''}, \textit{``1....*''}) of the multiplicity relationships that the pattern in Figure \ref{BlueROV2_ACP} depicts. This pattern contains the following multiplicity relationships: \textit{HasMultiplicity (S1, G3, 1 of *), HasMultiplicity (S4, A1, 0 of *), HasMultiplicity (G5, G10, 1 of *)}. According to these multiplicity rules, the ground-truth assurance case contains three instances of G3, three instances of A1, and one instance of G10.

However, as illustrated in Figure \ref{4Turbo_BlueROV2_AC}, GPT-4 Turbo generated only one instance of each element (G3, A1, and G10), which is fewer than specified in the ground-truth assurance case. On the other hand, GPT-4o generated three instances of each element (G3, A1, and G10), adhering more closely to the ground truth assurance case. Also, GPT-4o further developed the other two instances of G3 that the ground-truth assurance case left undeveloped. This may explain the higher number of GSN elements (i.e. 42 elements) that GPT-4o generated.

The discrepancy in the number of generated elements between both LLMs and the ground-truth assurance case can thus be partly explained by the ambiguity in the cardinality of the multiplicity relationships in the pattern illustrated in Figure \ref{BlueROV2_ACP}. We should point out that, occasionally, both LLMs might overlook removing the pattern decorators after instantiating a given pattern to create an assurance. Removing the pattern decorators is critical because their presence introduces ambiguity about whether each element in the generated assurance case is fully developed and instantiated. This ambiguity can lead to doubts about the completeness and reliability of the assurance case, potentially undermining its use for certification and compliance purposes.

In the remainder of this section, we further discuss the experiment results in light of the RQ2 four sub-research questions.

\subsubsection{RQ2.1: Comparative Analysis of One-Shot vs Zero-Shot Experiments}

The results reported in Table \ref{tab2} to \ref{tab4} show that when the median values are ranked from highest to lowest for a given system-model-metric combination across different experiments (i.e. Experiments \ref{exp:2} to \ref{exp:9}), the one-shot experiments (i.e. Experiments \ref{exp:2}, \ref{exp:4}, \ref{exp:6}, \ref{exp:8}) tend to achieve higher median values across various metrics compared to the zero-shot experiments (i.e. Experiments \ref{exp:3}, \ref{exp:5}, \ref{exp:7}, \ref{exp:9}).  Experiment \ref{exp:6}, in particular, consistently yields the highest metric values. 

Based on these results, we can conclude that the one-shot prompting approach is more effective for the automatic instantiation of assurance cases from assurance case patterns. Providing one example gives the LLMs a specific reference point to learn from, which aids in understanding the pattern instantiation task better and ensures that the generated assurance cases are closer to the ground truth assurance cases.

\subsubsection{RQ2.2: Impact of Domain Information}

The results presented in Tables \ref{tab2} to \ref{tab4} indicate that experiments incorporating domain information (i.e. Experiments \ref{exp:2}, \ref{exp:3}, \ref{exp:6}, \ref{exp:7}) yielded significantly higher median values across the three evaluation metrics compared to those without domain information (Experiments \ref{exp:4}, \ref{exp:5}, \ref{exp:8}, \ref{exp:9}). GPT-4o shows significant improvements when domain information is included in its prompts, thus achieving higher scores across all metrics. GPT-4 Turbo also benefits from this inclusion but to a lesser degree.

These findings highlight the crucial role of domain information in maintaining high performance across most metrics, especially semantic similarity, where the highest median semantic similarity values for different system-model combinations were observed in experiments with domain knowledge — with the only exception being Experiment \ref{exp:4} and \ref{exp:5} under GPT-4 Turbo for the ACAS XU system that is tied for highest with Experiment \ref{exp:2}.

\subsubsection{RQ2.3: Impact of Contextual Information}
\label{section-impact-context}


Experiments leveraging contextual information are Experiments \ref{exp:2}, \ref{exp:3}, \ref{exp:4}, and \ref{exp:5}. The results reported in Tables \ref{tab2} to \ref{tab4} indicate that experiments with and without context information performed relatively well. However, a notable distinction emerges when ranking the performance of experiments without context information (Experiments \ref{exp:6}, \ref{exp:7}, \ref{exp:8}, \ref{exp:9}). Experiments \ref{exp:8} and \ref{exp:9}, which also lack domain information, often yield the lowest median of metric values compared to other experiments for a given system-model-metric combination. Interestingly, when counting the frequency or number of times an experiment performs the best or gives the highest results across all system-model-metric combinations, Experiment \ref{exp:6} (one-shot, without context information, with domain information, and with predicate rules) emerges with the highest frequency of top values across different metrics, models, and systems.

Initially, we expected Experiment \ref{exp:2} (one-shot, with context information, with domain information, and with predicate rules) to yield the highest frequency of top values and be the best experiment overall. This expectation was based on the assumption that having comprehensive context information would significantly enhance the model's performance. However, the results indicate otherwise. Experiment 6 outperformed all other experiments, including Experiment 2, consistently achieving the highest frequency of top metric values.

One possible reason for this outcome could be our predicate-based rules, formulated in accordance with the guidelines, elements, and decorators outlined in the GSN Standard. These GSN characteristics also informed our context information. The use of predicate-based rules that comply with the GSN standard may have contributed to the success of Experiment \ref{exp:6}, even in the absence of explicit context information. Additionally, it is possible that both GPT-4o and GPT-4 Turbo, have a prior understanding of the information embedded within our context information, and hence, were able to perform effectively despite the lack of explicit context information in Experiment \ref{exp:6}.

In conclusion, although incorporating context information was anticipated to enhance performance, its absence, particularly when combined with a lack of domain information, predicate-based rules, and one-shot examples, can lead to poorer results. While context information alone does not drastically affect performance, the presence of predicate-based rules, domain information, and one-shot examples significantly mitigates the negative impact of missing context information.

\subsubsection{RQ2.4: Impact of Predicate-based Rules}

Recall from above that, in each of the eight experiments with SE knowledge, we specify the predicate-based rules in LLMs prompts. The results presented in Tables \ref{tab2} to \ref{tab4} indicate that Experiment \ref{exp:9} (zero-shot, without context information, without domain information, and with predicate-based rules) frequently yields the lowest median values compared to all other experiments when these median values are ranked from highest to lowest for a given system-model-metric combination. This suggests that predicate-based rules are not meant to be utilized alone for the automatic instantiation of assurance case patterns. Rather, they should be used in combination with other categories of SE knowledge (e.g., domain information, context information, and one-shot examples) to ensure the generation of assurance cases that are close to the ground truth.

\smallskip
\noindent\fbox{%
    \parbox{\linewidth}{%
    \smallskip
In summary, for \textbf{RQ2}, our experiments demonstrate that incorporating SE knowledge into LLMs significantly enhances their ability to generate assurance cases that comply with given patterns and closely align with ground-truth assurance cases. Experiment 6, which leverages a one-shot example, domain information, and predicate-based rules, consistently showed superior performance compared to other experiments. For instance, it achieved a median exact match value of 0.81 for the BlueROV2 system and a median BLEU score of 0.76 for the ACAS XU system.

Cardinality ambiguity in multiplicity relationships posed challenges, particularly for the GPCA and IM Software systems, resulting in lower exact match and BLEU score values across all experiments, with the highest median exact match for both systems being equal to 0.35 and 0.28 respectively. However, domain information significantly improved performance, as evidenced in experiments incorporating it, where models achieved higher semantic similarity scores. For example, ACAS XU under GPT-4o achieved a score of 0.92 in Experiments \ref{exp:2}, \ref{exp:3}, and \ref{exp:6}, while GPCA saw a score of 0.81 in Experiment \ref{exp:2} under GPT-4o, and IM Software achieved 0.71 in Experiments \ref{exp:6} and \ref{exp:7}.

Notably, semantic similarity scores were generally higher compared to Exact match and BLEU scores across most experiments, indicating that while text wordings might differ, the generated assurance cases were still meaningfully close to the ground-truth assurance cases. Comparative analysis also revealed that the one-shot approach generally produced better results than the zero-shot method.

Also, while we initially expected that having comprehensive context information would significantly enhance model performance, experiments without explicit context information but with predicate-based rules, domain information, and one-shot examples still performed exceptionally well. Experiment \ref{exp:6}, which lacked context information yet incorporated these other categories of SE knowledge, frequently outperformed experiments that included context information.

Finally, incorporating various categories of SE knowledge helps achieve reliable and effective LLM-generated assurance cases that comply with given patterns, while addressing complexities like ambiguous cardinality in patterns further ensures that LLM-generated assurance cases closely align with ground-truth cases.

 }}

\subsection{RQ3:  Which of the evaluated LLMs performs best when it comes to automatically instantiating assurance cases from assurance case patterns?}

Tables \ref{tab2} to \ref{tab4} highlight in bold the higher median value for a given system-model combination under the different metrics. Evaluating which model has the higher median value for each system-model combination across the different metrics and experiments with SE knowledge shows that GPT-4o achieves 77\% of the highest median scores across the various metrics and experiments for the different systems. On the other hand, GPT-4 Turbo achieves 17\% of the highest median scores across the various metrics and experiments for the different systems. The remaining 6\% represents the single instance where both GPT-4o and GPT-4 Turbo are tied for having the same highest median value in an experiment across the various metrics for the different systems. Based on this analysis, GPT-4o outperforms GPT-4 Turbo in the median values for the Exact Match, BLEU score, and semantic similarity results across the majority of our experiments for each system. This suggests that GPT-4o seems to understand subtle ACP-related nuances like cardinality ambiguity in multiplicity relationships better than GPT-4 Turbo.

\smallskip
\noindent\fbox{%
    \parbox{\linewidth}{%
    \smallskip
   In summary, for \textbf{RQ3}, our analysis reveals that GPT-4o outperforms GPT-4 Turbo in 77\% of the 8 experiments incorporating SE knowledge, across various metrics and systems. Overall, GPT-4o shows an enhanced grasp of SE knowledge, including nuanced pattern details like cardinality ambiguity in multiplicity relationships, enabling it to produce assurance cases that are closer to the ground truth compared to GPT-4 Turbo.
    }}

 
\section{Discussion}
\label{sec:Discussion}

\subsection{Analyzing varying One-Shot example}

As stated in Section \ref{Section:deepmind}, to conduct our four one-shot experiments with SE knowledge, we randomly picked DeepMind's AC together with its ACP as an example. Still, in this section, we aim to determine if picking a specific example 
over another when running one-shot experiments has a significant impact on the quality of the results. For brevity's sake, we only focus in this section on Experiment \ref{exp:2} i.e. the experiment that leverages all the combinations of SE knowledge. 

\subsubsection{Methodology}
To assess the impact of varying the example on Experiment \ref{exp:2} results, we adapted the popular Leave One Out Cross-Validation (LOOCV) \citep{b134, b135} method to split our example and test data. This ensures the utilization of all systems in our dataset for both the one-shot example and testing data. Given that the number of systems (\textit{N}) in our dataset is 5, we iterated and picked one system from \textit{N} to use as a one-shot example while we used the remaining \textit{N-1} systems as test data to validate the LLMs performance. We iterated over \textit{N} until all systems in \textit{N} have been utilized as a one-shot example. Besides, we ran Experiment \ref{exp:2}, \textit{K=5} times, and we calculated the median metric values across the 5 runs for the different one-shot examples. 

\subsubsection{Results}

Tables \ref{tab5}, \ref{tab6}, and \ref{tab7} report the comparison of the median metric results the LLMs yield when we vary the one-shot example in Experiment \ref{exp:2} across five runs. Each column under the \textit{``One-shot Example''} header specifies each system used as a one-shot example. Each row under the \textit{``Test System''} header represents each system used as test data. \textit{``Null''} values indicate cases where the same system is meant to serve as both the test and one-shot example, which we do not evaluate. The values highlighted in bold indicate the highest median value for a given test system across the different one-shot examples. 

\begin{table} [h!]
\centering

\caption{Exact Match Comparison of Varying One-Shot Example}
\label{tab5}
\begin{tabular}{|l |l |c  |c |c |c |c|} \hline       
\multirow{2}{*}{\textbf{Test System}} & \multirow{2}{*}{\textbf{Model}} & \multicolumn{5}{|c|}{\textbf{One-Shot Example}} \\  \cline{3-7}
 &  & \textbf{ACAS XU} & \textbf{BlueROV2} & \textbf{DeepMind} & \textbf{GPCA} & \textbf{IM Software} \\ \hline 
\multirow{2}{*}{ACAS XU} & GPT-4o & Null &\textbf{ 0.87} & 0.79 & 0.78 & 0.75 \\ \cline{2-7}
 & GPT-4 Turbo & Null & 0.6 & 0.65 &\textbf{ 0.66} & 0.5 \\ \hline 
\multirow{2}{*}{BlueROV2} & GPT-4o & 0.2 & Null & \textbf{0.65} & 0.2 & 0.21 \\ \cline{2-7}
 & GPT-4 Turbo & \textbf{0.78} & Null & 0.76 & 0.72 & 0.75 \\ \hline 
\multirow{2}{*}{DeepMind} & GPT-4o & 0.3 & \textbf{0.35} & Null & 0.27 & 0.29 \\ \cline{2-7}
 & GPT-4 Turbo & 0.25 & \textbf{0.3} & Null & 0.28 & 0.23 \\ \hline 
\multirow{2}{*}{GPCA} & GPT-4o & \textbf{0.31} & 0.22 & 0.26 & Null & 0.27 \\ \cline{2-7}
 & GPT-4 Turbo & \textbf{0.28} & \textbf{0.28} & 0.25 & Null & 0.24 \\ \hline 
\multirow{2}{*}{IM Software} & GPT-4o & \textbf{0.36} & 0.28 & 0.1 & 0.29 & Null \\ \cline{2-7}
 & GPT-4 Turbo & 0.09 & 0.17 & 0.09 & \textbf{0.25} & Null \\ \hline 

\end{tabular}

\end{table}

\begin{table} [h!]
\centering

\caption{BLEU Score Comparison of Varying One-Shot Example}
\label{tab6}
\begin{tabular}{|l |l |c  |c |c |c |c|} \hline       
\multirow{2}{*}{\textbf{Test System}} & \multirow{2}{*}{\textbf{Model}} & \multicolumn{5}{|c|}{\textbf{One-Shot Example}} \\ \cline{3-7}
 &  & \textbf{ACAS XU} & \textbf{BlueROV2} & \textbf{DeepMind} & \textbf{GPCA} & \textbf{IM Software }\\ \hline 
\multirow{2}{*}{ACAS XU} & GPT-4o & Null & 0.76 & 0.73 & \textbf{0.81} & 0.74 \\ \cline{2-7}
 & GPT-4 Turbo & Null & 0.7 & \textbf{0.71} & 0.69 & 0.58 \\ \hline 
\multirow{2}{*}{BlueROV2} & GPT-4o & 0.44 & Null &\textbf{ 0.54} & 0.4 & 0.4 \\ \cline{2-7}
 & GPT-4 Turbo & 0.6 & Null & 0.58 & \textbf{0.63} & 0.58 \\ \hline 
\multirow{2}{*}{DeepMind} & GPT-4o & 0.21 & \textbf{0.22} & Null & 0.17 & 0.21 \\ \cline{2-7}
 & GPT-4 Turbo & 0.16 & 0.19 & Null & 0.2 & \textbf{0.22} \\ \hline 
\multirow{2}{*}{GPCA} & GPT-4o & 0.26 & 0.23 & \textbf{0.32} & Null & 0.29 \\ \cline{2-7}
 & GPT-4 Turbo & 0.24 & 0.28 & 0.16 & Null & \textbf{0.3} \\ \hline 
\multirow{2}{*}{IM Software} & GPT-4o & 0.29 & \textbf{0.31} & 0.27 & 0.29 & Null \\ \cline{2-7}
 & GPT-4 Turbo & 0.19 & \textbf{0.31} & 0.17 & \textbf{0.31} & Null \\ \hline 

\end{tabular}

\end{table}

\begin{table}
\centering

\caption{Semantic Similarity Comparison of Varying One-Shot Example}
\label{tab7}
\begin{tabular}{|l |l |c  |c |c |c |c|} \hline       
\multirow{2}{*}{\textbf{Test System}} & \multirow{2}{*}{\textbf{Model}} & \multicolumn{5}{|c|}{\textbf{One-Shot Example}} \\ \cline{3-7}
 &  & \textbf{ACAS XU} & \textbf{BlueROV2} & \textbf{DeepMind} & \textbf{GPCA} & \textbf{IM Software} \\ \hline 
\multirow{2}{*}{ACAS XU} & GPT-4o & Null & \textbf{0.93} & 0.92 & \textbf{0.93} & 0.9 \\ \cline{2-7}
 & GPT-4 Turbo & Null & 0.9 & \textbf{0.91} & 0.9 & 0.86 \\ \hline 
\multirow{2}{*}{BlueROV2} & GPT-4o & \textbf{0.89} & Null & 0.88 & 0.86 & \textbf{0.89} \\  \cline{2-7}
 & GPT-4 Turbo & 0.86 & Null & \textbf{0.89} & 0.85 & 0.85 \\ \hline 
\multirow{2}{*}{DeepMind} & GPT-4o & 0.59 & 0.59 & Null & \textbf{0.61} & \textbf{0.61} \\  \cline{2-7}
 & GPT-4 Turbo & 0.54 & 0.52 & Null & 0.55 & \textbf{0.59} \\ \hline 
\multirow{2}{*}{GPCA} & GPT-4o & 0.59 & 0.74 & \textbf{0.81} & Null & 0.67 \\  \cline{2-7}
 & GPT-4 Turbo & 0.55 & 0.58 & 0.56 & Null & \textbf{0.59} \\ \hline 
\multirow{2}{*}{IM Software} & GPT-4o & \textbf{0.71} & 0.7 & 0.7 & \textbf{0.71} & Null \\  \cline{2-7}
 & GPT-4 Turbo & 0.65 & 0.69 & 0.65 & \textbf{0.71} & Null \\ \hline 

\end{tabular}

\end{table}

The results in Tables \ref{tab5} to \ref{tab7} indicate that when analyzing the frequency of the top metric results across the various metrics and test systems, the ACs and ACPs of both GPCA and DeepMind might be the most effective as one-shot examples. For instance, when including counts of ties, the GPCA system emerges as the best one-shot example while when excluding the six tie counts, the frequencies shift, making DeepMind the best one-shot example.

Tables \ref{tab5} to \ref{tab7} also indicate that using the AC and ACP of either ACAS Xu or IM Software as a one-shot example produces the lowest count of top values across all metrics. Specifically, the IM Software as an example
fails to achieve any top values under the Exact Match metric while the ACAS Xu system as an example fails to achieve any top values under the BLEU score metric. This indicates that, among all the systems in our dataset, both the IM Software system and ACAS Xu system might be the least effective as a one-shot example.

\smallskip
\noindent\fbox{%
    \parbox{\linewidth}{%
    \smallskip
    Our analyses of the results we obtained when varying different systems' ACs and ACPs as one-shot examples highlight that the selection of a specific system's AC and ACP as an example is crucial, as it influences the performance of LLMs across different metrics. Thus, choosing the most suitable one-shot example can significantly improve the quality of one-shot experiments.
    }}

\subsection{Are Human Experts Still Needed for Assurance Case Creation in the Age of LLMs?} 

We have manually assessed the best results the two LLMs produced to determine if they are equivalent to the ones generated by human experts and more specifically to the ones assurance case developers usually create. We further discuss that work below.

\subsubsection{Methodology used for the manual assessment}
As we stated in Section \ref{section-impact-context}, Experiment \ref{exp:6} yields the best results for both LLMs. We therefore decided to focus on the manual assessment of that experiment results. To manually assess these results, we relied on a metric called \textbf{\textit{reasonability}} \citep{b74, b107, b114, b130}. A reasonable GSN element is a GSN element that "\textit{could reasonably be in the ground-truth but is not}" \citep{b107, b130}. The reasonability metric allows assessing the degree to which the assurance cases generated by LLMs are useful, coherent, and contain GSN elements that are valid but that are not present in the ground-truths i.e. GSN elements which the human experts have not thought of but that could have enriched the assurance case. 

Two researchers (i.e. raters) --and more specifically two authors with strong experience in system assurance and GSN --  independently assessed the reasonability of the forty assurance cases the two LLMs collectively generated for Experiment \ref{exp:6}. These researchers utilized a linear scale to assess the reasonability of each of the corresponding assurance cases. In that scale, 1 equals \textit{Totally reasonable}, 2 signifies \textit{Mostly reasonable}, 3 means \textit{Moderately reasonable}, 4 represents \textit{Slightly reasonable}, and 5 denotes \textit{Unreasonable}. To assess the inter-rater reliability, we relied on Kendall's Tau \citep{b138}
 as in the literature (e.g., \citep{b97, b107, b74}). Kendall's Tau is a correlation coefficient that varies between - 1 and 1. A value equal to 1 indicates a strong level of agreement between raters. A negative value indicates there is no agreement between raters. As in \citep{b97, b107}, we relied on 
an online tool called \textbf{\textit{GIGA calculator}} \footnote{https://www.gigacalculator.com/calculators/correlation-coefficient-calculator.php} to automatically compute the value of that coefficient with a 95\% confidence interval.

Note that to help in the automatic conversion of the assurance cases Experiment 6 yields, we relied on \textit{SmartGSN}\footnote{SmartGSN is the prototype of a web-based tool whose main features include  converting assurance cases from structured prose to GSN diagrams. The core technologies used to develop \textit{SmartGSN} are ReactJS and Google FireBase. More specifically, \textit{SmartGSN} relies on React Flow, a  customizable React component, for the implementation of its node-based editors and interactive diagrams. }.

\subsubsection{Discussion of the reasonability results}
The Kendall's Tau value obtained from reasonability ratings is equal to \textbf{0.69} when considering all the ratings the two raters provided for both the assurance cases that both GPT-4o and GPT-4 Turbo generated. This indicates a good agreement between the two researchers who graded Experiment \ref{exp:6} results. This means that the two researchers who graded Experiment \ref{exp:6}  results produced consistent, reliable and similar ratings.

Table \ref{tab8} reports for each rater, for each LLM and for each system, the average of the reasonability ratings. That Table also aggregates these ratings for each LLM at hand. Hence, the averages of the reasonability scores across all the forty assurance cases Experiment \ref{exp:6} yields and across the two raters are respectively  2.75  for GPT-4o and 3.05 for GPT-4 Turbo. These averages are close to 3 i.e. Moderalely reasonable. Thus, the reasonability results demonstrate that LLMs like GPT-4o and GPT-4 Turbo do quite well at generating assurance cases from assurance case patterns. Particularly in one-shot settings such as Experiment \ref{exp:6}, where domain information and predicate-based rules are leveraged, these models showcase remarkable proficiency in assurance case generation.

These findings suggest that LLMs can effectively perform a significant portion of the pattern instantiation process, offering invaluable time and resource savings, especially in generating initial drafts of assurance cases — a task that can be laborious, error-prone, and time-consuming when performed manually \citep{b88, b120}.

However, despite the performance of LLMs in assurance case generation, human expertise remains currently indispensable. Assurance cases often demand adherence to specific standards or regulatory requirements, as well as an understanding of subtle pattern-related nuances (e.g., an inferred cardinality for a multiplicity relationship) that LLMs may not fully grasp or overlook. There's also the potential for LLMs to hallucinate, especially when complex context and domain information are involved, raising concerns about the reliability of generated assurance cases.

To address these limitations, we believe a semi-automatic approach may be more suitable to create assurance cases. In this regard, one can leverage LLMs for the automatic instantiation of assurance case patterns to create initial assurance case drafts. Subsequently, human experts (i.e. assurance case developers) can refine and adjust these drafts, ensuring they meet necessary standards, address potential gaps or inconsistencies, and enhance the overall quality of the assurance process. This is in accordance with \citet{b74} who concluded that LLMs are still not able to fully automate the domain modeling task. However, a human modeler can continuously provide feedback to the model to improve the model's output incrementally. This also aligns with \citet{b130} who concluded that while LLMs can significantly accelerate the development of safety cases, the expertise and oversight of human safety case developers remain indispensable, particularly for ensuring the highest levels of safety assurance.

\begin{table}
\centering
\caption{Average Reasonability Rating of Assurance Cases Experiment \ref{exp:6} yields}
\label{tab8}
\begin{tabular}{|l |l |l  |l|} \hline     
\multirow{2}{*}{\textbf{Model}} & \multirow{2}{*}{\textbf{System}} & \multicolumn{2}{|l|}{\textbf{Average Reasonability Ratings}} \\ \cline{3-4}
 &  & \textbf{Rater 1} & \textbf{Rater 2} \\ \hline 
\multirow{4}{*}{GPT-4o} & ACAS XU & 2 & 2 \\ \cline{2-4}
 & BLUEROV2 & 3.2 & 2.4 \\ \cline{2-4}
 & GPCA & 3.8 & 3.2 \\ \cline{2-4}
 & IM SOFTWARE & 3 & 2.4 \\ \hline 
\multirow{4}{*}{GPT-4 Turbo} & ACAS XU & 2.2 & 2.4 \\ \cline{2-4}
 & BLUEROV2 & 3.4 & 2.6 \\ \cline{2-4}
 & GPCA & 3.6 & 3 \\ \cline{2-4}
 & IM SOFTWARE & 3.8 & 3.4 \\ \hline 

\end{tabular}

\end{table}

\section{Threats to Validity}
\label{sec:Validity_Threats}

\subsection{Internal Validity}
The dataset used in our experiment consists of six assurance case patterns and five partial assurance cases complying with these patterns. We experienced difficulties in obtaining full assurance cases complying with a given pattern due to the large size of these documents and the sensitive and confidential nature of the information contained in them \citep{b121}. This limits the availability of full assurance case patterns and derived assurance cases from these patterns, as they are not readily published or available. To mitigate this, we selected patterns and assurance cases spanning various application domains. We also contacted some of these assurance case developers, which was usually fruitless. 

The threat identification pattern that Figure \ref{ACAS_XU_ACP} depicts, and that is part of our dataset, is divided into two parts. The first part is highlighted in red while the second part is highlighted in blue. One potential threat to the validity of our work may emerge from the ground-truth assurance case derived from this pattern. \citet{b126} in their study, provided only an instantiation of the part highlighted in blue due to brevity's sake. They stated that the placeholder \textit{``{System}''} in the uninstantiated nodes (C0, G0, G1, G2) in the part highlighted in red would be replaced by ``ACAS Xu''. To obtain a complete assurance case, we refined their assurance case. Hence, we manually instantiated the elements in the initial part (highlighted in red) of the threat identification pattern by simply replacing the placeholder \textit{``{System}''} by "ACAS Xu".
This manual instantiation of the initial part (elements C0, G0, G1, G2) may pose a validity threat, as it may introduce a potential human error. The latter could impact the consistency and correctness of the so-obtained ground-truth assurance case. This underscores the need for methods that facilitate the automatic instantiation of assurance cases from patterns.

To mitigate the aforementioned threats in future work, we aim to partner with industry to gain access to full assurance cases derived from a given pattern. 

\subsection{Construct Validity}
We extracted the domain information used in our experiments from the following cited references which describe our dataset: \citep{b126, b88, b9, b11}. Thus, the ACs generated by the two LLMs at hand relied solely on the information available in the cited references. Consequently, key details such as arguments or artifacts related to real-life data or scenarios might have been omitted when deriving domain information. It is also possible that subtle details of the way we formalized the patterns based on the GSN standard might have influenced the results. Future work could try repeating the experiments with small permutations on the way the patterns are expressed.

Also, one potential threat to the validity of our results is the number of runs (K = 5) used in our experiments. By performing each experiment five times, we aimed to capture a sufficient amount of variability and ensure the reliability of our findings. We picked that number in accordance with the literature focusing on the use of LLMs to automate software modeling tasks (e.g., \cite{b74, b107}).  However, this number of runs may not fully reflect the variability in the non-deterministic results generated by our LLMs. To mitigate this threat and enhance the robustness of our findings, we utilized various test systems across different domains and included varied experimental conditions, categorizing different types of SE knowledge. However, in future work, we aim to increase the number of runs to ensure greater confidence in our findings.

\subsection{Conclusion Validity}

The knowledge cut-off date for our two LLMs is 2023. The dataset utilized in our work was published before this date, suggesting that our models' training data might overlap with our dataset, potentially affecting the generalizability of our results. To address this in accordance with \citet{b114}, we formalized our assurance case patterns in the predicated-based format to obtain representations that both LLMs have never seen before. Still, in future work, we plan to validate the effectiveness of our approach using more recent datasets coming from the industry. These ones are not publicly available.

\section{Conclusion and Future Work}
\label{sec:Conclusion}
In this study, we relied on large language models (LLMs) to support the automatic instantiation of assurance cases in compliance with assurance case patterns formalized using predicate-based rules. We conducted a variety of experiments across four systems to evaluate the impact of different categories of software engineering knowledge on the performance the LLMs yield when instantiating assurance cases. 

Our experiment results show large language models can automatically generate relatively good assurance cases when leveraging software engineering knowledge, including knowledge represented in the form of patterns. Still, our experiments also show that we still need human expertise to refine the LLM-generated assurance cases to make them suitable for the certification of mission-critical systems. This is in accordance with other results in the literature (e.g., \citep{b74, b107}).

In future work, we will further validate our approach by conducting more experiments aiming at instantiating additional assurance case patterns using a wider range of LLMs.  
Since our approach can be applied to other contexts that may be very different, we also plan to explore the use of our approach to instantiate software artifacts. Examples could include generating software designs from design patterns, or software architecture models from architectural patterns. 

\section*{Acknowledgement}
We would like to thank the Mitacs Globalink Research Internship program for helping us secure the funding required to develop \textit{SmartGSN}. We would like to thank the two Mitacs interns (i.e. 
Emiliano Berrones Gutiérrez and Daniel Méndez Beltrán) who are currently developing  \textit{SmartGSN}. We would also like to thank two former graduate students (i.e. Mithila Sivakumar and Kimya Khakzad Shahandashti) for their involvement in the early development of a former \textit{SmartGSN} version.

\appendix
\section{Appendix}
\label{Appendix}

\subsection{Contextual Information} 
\label{appendix:contextual-information}
 ~\\
@Context\_AC\newline 

An assurance case, such as a safety case or security case, can be represented using Goal Structuring Notation (GSN), a visual representation that presents the elements of an assurance case in a tree structure. The main elements of a GSN assurance case include Goals, Strategies, Solutions (evidence), Contexts, Assumptions, and Justifications. 

Additionally, an assurance case in GSN may include an undeveloped element decorator, represented as a hollow diamond placed at the bottom center of a goal or strategy element. This indicates that a particular line of argument for the goal or strategy has not been fully developed and needs to be further developed.

I will explain each element of an assurance case in GSN so you can generate it efficiently.

\begin{enumerate}
    \item Goal – A goal is represented by a rectangle and denoted as G. It represents the claims made in the argument. Goals should contain only claims. For the top-level claim, it should contain the most fundamental objective of the entire assurance case.

    \item Strategy – A strategy is represented by a parallelogram and denoted as S. It describes the reasoning that connects the parent goals and their supporting goals. A Strategy should only summarize the argument approach. The text in a strategy element is usually preceded by phrases such as “Argument by appeal to…”, “Argument by …”, “Argument across …” etc.

    \item Solution – A solution is represented by a circle and denoted as Sn. A solution element makes no claims but are simply references to evidence that provides support to a claim.

    \item Context (Rounded rectangles) – In GSN, context is represented by a rounded rectangle and denoted as C. The context element provides additional background information for an argument and the scope for a goal or strategy within an assurance case.

    \item Assumption – An assumption element is represented by an oval with the letter ‘A’ at the top- or bottom-right. It presents an intentionally unsubstantiated statement accepted as true within an assurance case. It is denoted by A

    \item Justification (Ovals) – A justification element is represented by an oval with the letter ‘J’ at the top- or bottom-right. It presents a statement of reasoning or rationale within an assurance case. It is denoted by J.
\end{enumerate}

@End\_Context\_AC
\newline

@Context\_ACP
\newline

Assurance case patterns in GSN (Goal Structuring Notation) are templates that can be re-used to create an assurance case. Assurance case patterns encapsulate common structures of argumentation that have been found effective for addressing recurrent safety, reliability, or security concerns. An assurance case pattern can be instantiated to develop an assurance case by replacing generic information in placeholder decorator with concrete or system specific information.

To represent assurance case patterns in GSN format, additional decorators have been provided to support assurance case patterns. These additional decorators are used together with the elements of an assurance case to represent assurance case pattern. I will explain each additional decorator below to support assurance case pattern in GSN.

\begin{enumerate}
    \item Uninstantiated - This decorator denotes that a GSN element remains to be instantiated, i.e. at some later stage, the generic information in placeholders within a GSN element needs to be replaced (instantiated) with a more concrete or system specific information. This decorator can be applied to any GSN element.

    \item Uninstantiated and Undeveloped – Both decorators of undeveloped and uninstantiated are overlaid to form this decorator. This decorator denotes that a GSN element requires both further development and instantiation. 

    \item Placeholders – This is represented as curly brackets “{}” within the description of an element to allow for customization. The placeholder "{}" should be directly inserted within the description of elements for which the predicate "HasPlaceholder (X)" returns true. The placeholder "{}" can sometimes be empty or contain generic information that will need to be replaced when an assurance case pattern is instantiated. 

    \item Choice - A solid diamond is the symbol for Choice. A GSN choice can be used to denote alternatives in satisfying a relationship or represent alternative lines of argument used to support a particular goal.

    \item Multiplicity - A solid ball is the symbol for multiple instantiations. It represents generalized n-ary relationships between GSN elements. Multiplicity symbols can be used to describe how many instances of one element-type relate to another element.

    \item Optionality - A hollow ball indicates ‘optional’ instantiation. Optionality represents optional and alternative relationships between GSN elements.

\end{enumerate}

The following steps is used to create an assurance case from an Assurance cases pattern.

\begin{enumerate}
    \item Create the assurance case using only elements and decorators defined for assurance cases.

    \item Remove all additional assurance case pattern decorators such as (Uninstantiated, Placeholders, Choice, Multiplicity, Optionality, and the combined Uninstantiated and Undeveloped decorator)

    \item Remove the placeholder symbol "{}" and replace all generic information in placeholders “{}” with system specific or concrete information. \newline

@End\_Context\_ACP

\end{enumerate}

\subsection{Assurance Case Pattern for ACAS XU System}
 \begin{figure} [h!]
   
    \centering
    \includegraphics[width=1\linewidth]{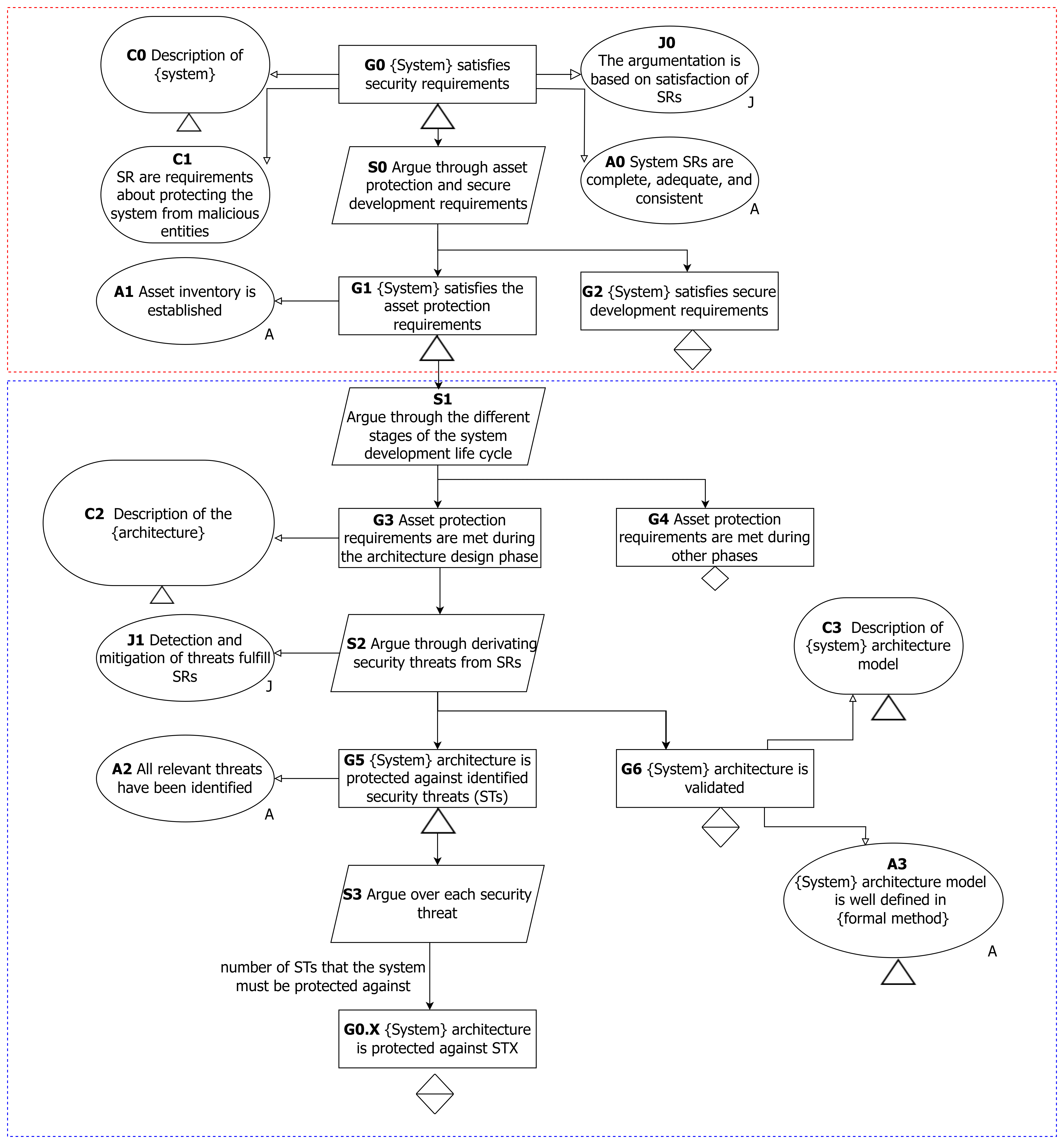}
    \caption{Assurance case pattern for Threat Identification - adapted from \citep{b126}}
    \label{ACAS_XU_ACP}
\end{figure}

\subsection{Assurance case for the BLUEROV2 System Generated by GPT-4o}
\begin{landscape}
\begin{figure} [h!]
    \centering
    \includegraphics[width=0.9\linewidth]{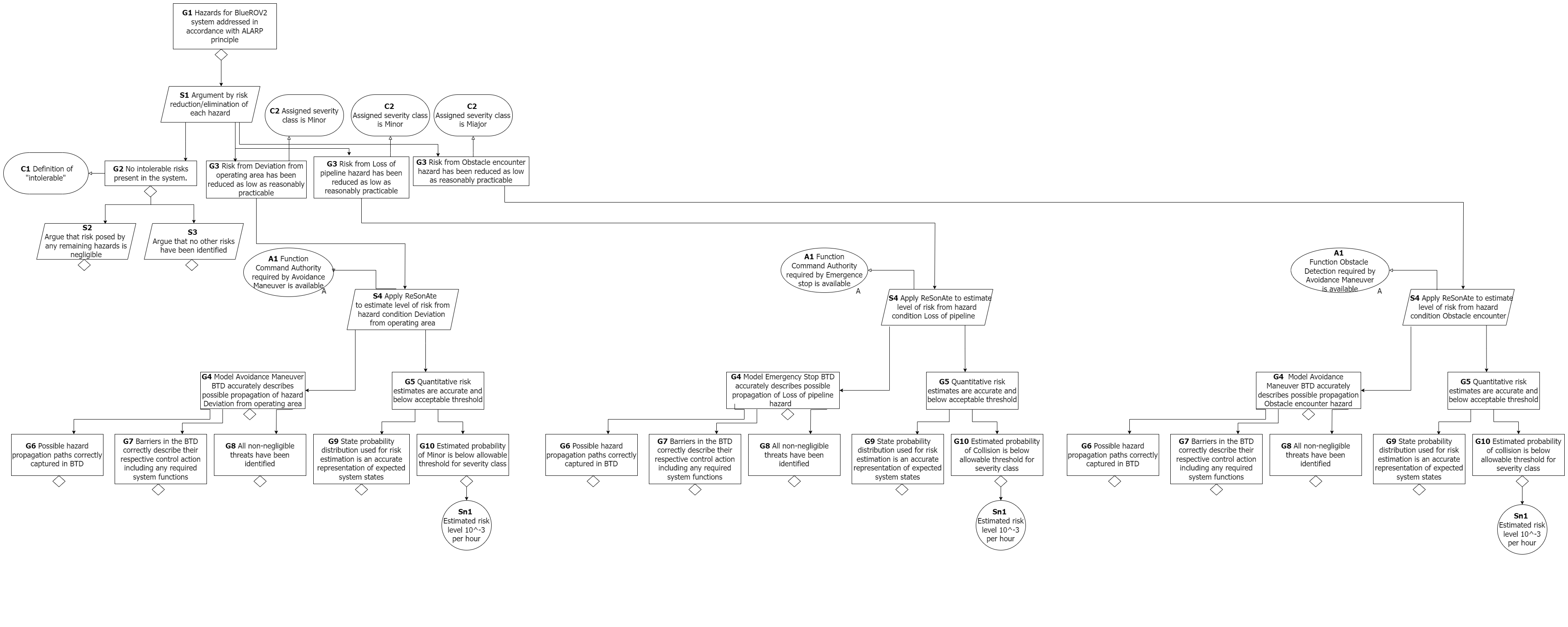}
    \caption{An Assurance case for the BLUEROV2 System Generated by GPT-4o} 
    \label{4o_BlueROV2_AC}
\end{figure}
\end{landscape}

\subsection{Assurance case for the BLUEROV2 System Generated by GPT-4 Turbo}
\begin{figure} [h!]
    \centering
    \includegraphics[width=1\linewidth]{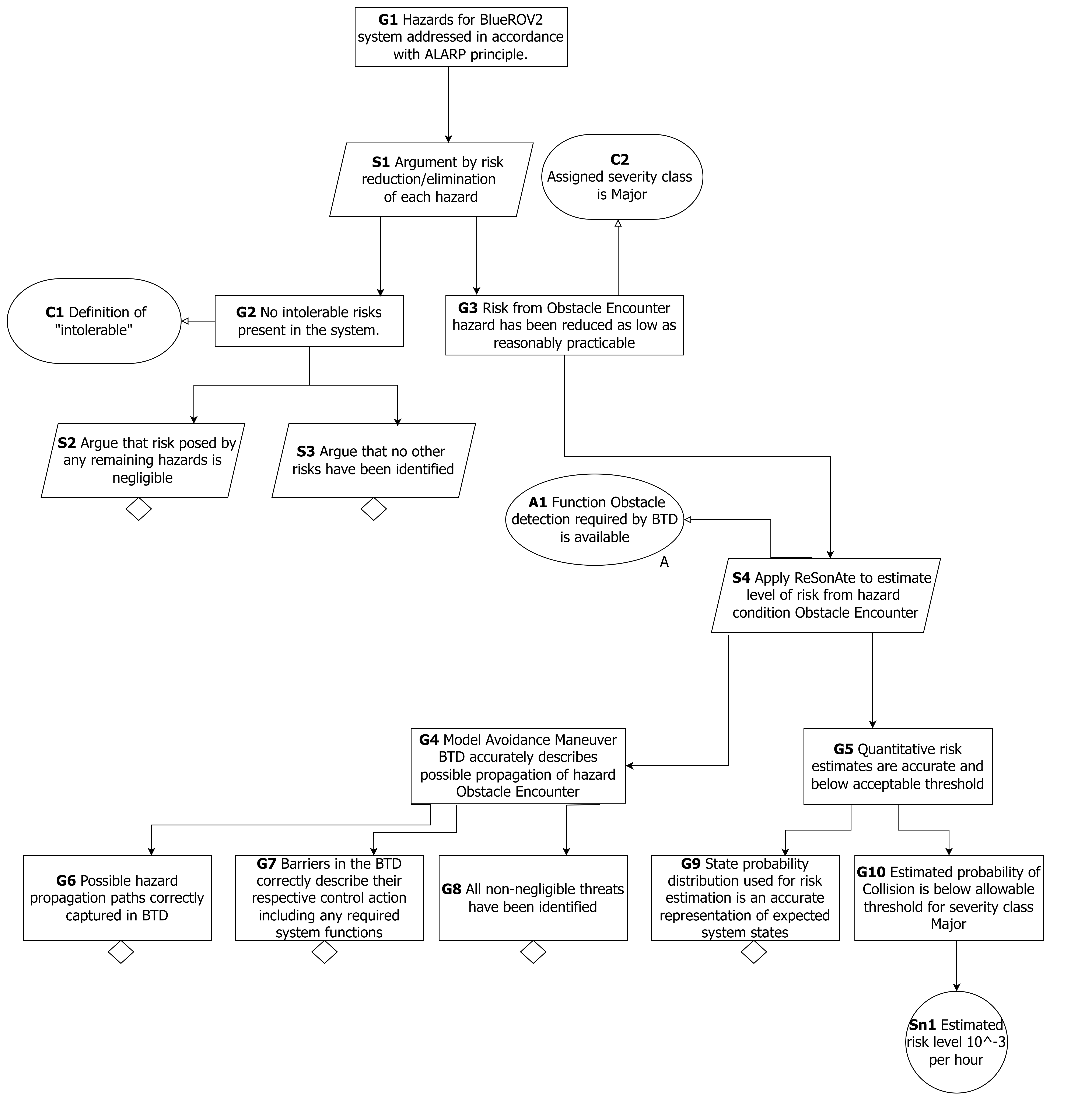}
   \caption{An Assurance case for the BLUEROV2 System Generated by GPT-4 Turbo}
    \label{4Turbo_BlueROV2_AC}
\end{figure}

\subsection{Assurance Case Pattern for BlueROV2 System}
\begin{figure} [h!]
    \centering
    \includegraphics[width=1\linewidth]{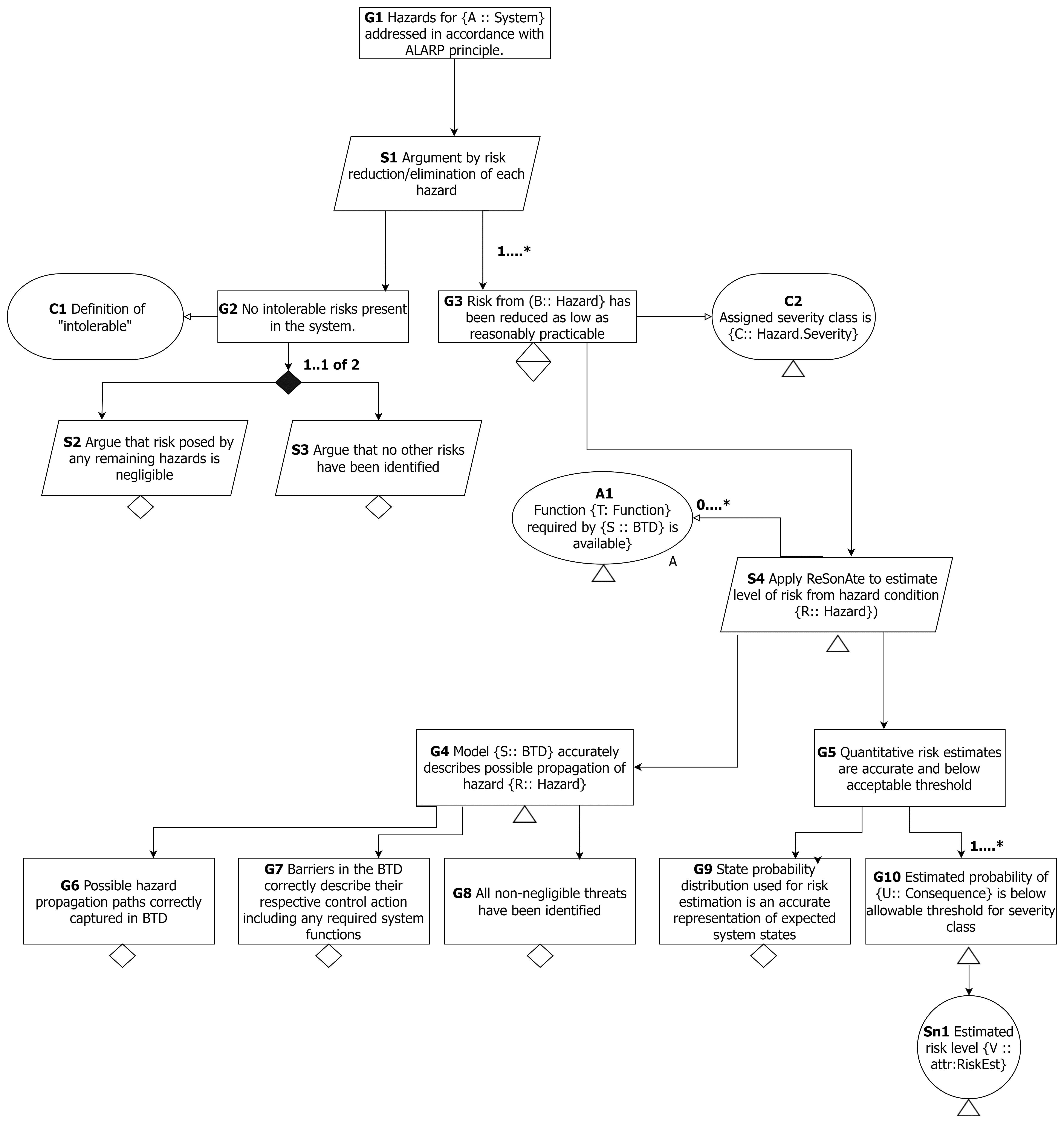}
    \caption{The ALARP pattern sequentially composed with the ReSonAte pattern - adapted from \citep{b88}} 
    \label{BlueROV2_ACP}
  
\end{figure}

\subsection{Ground Truth Assurance Case for the BlueROV2 System}

\begin{figure} [h!]
  
    \centering
    \includegraphics[width=1\linewidth]{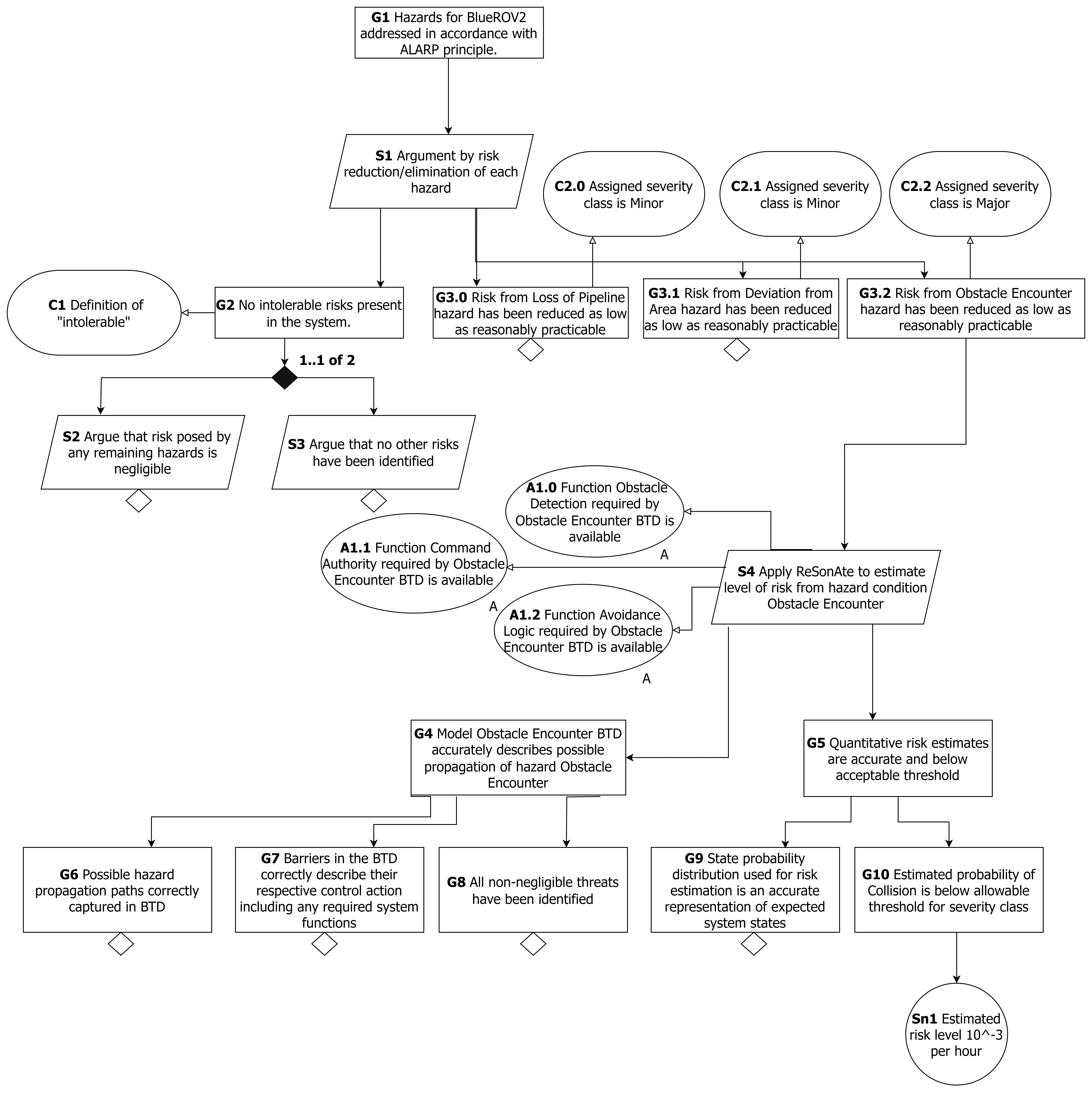}
    \caption{Ground Truth Assurance Case for the BlueROV2 System - adapted from \citep{b88}} 
  \label{BlueROV2_AC}
\end{figure}

\clearpage


\bibliographystyle{cas-model2-names}

\bibliography{mybib}



\end{document}